\documentclass[10pt,twoside]{article}

\usepackage{graphicx}
\usepackage{amssymb}
\usepackage{latexsym}

\makeatletter

\@addtoreset{equation}{section} \makeatother
\newtheorem{theorem}{Theorem}[section]

\newtheorem{lemma}[theorem]{Lemma}
\newtheorem{proposition}[theorem]{Proposition}
\newtheorem{corollary}[theorem]{Corollary}
\newtheorem{definition}[theorem]{Definition}
\newtheorem{example}[theorem]{Example}

\newcommand{\cvd}{\ \rule{0.5em}{0.5em}}
\newcommand{\be}{\begin{equation}}
\newcommand{\ee}{\end{equation}}

\font\ddpp=msbm10 scaled \magstep 1
\def\L{\hbox{\ddpp L}}
\newcommand{\N}{{\mathbb N}}

\newcommand{\R}{{\mathbb R}}

\title{\vspace{0.5in} {\bf  The Causal Boundary of Spacetimes Revisited}}

\author{ Jos\'e L. Flores
\\
{\it\small Departamento de \'Algebra, Geometr\'{\i}a y Topolog\'{\i}a}\\
{\it\small Facultad de Ciencias, Universidad de M\'alaga}\\
{\it\small Campus Teatinos, 29071 M\'alaga, Spain} \\ {\it\small
e-mail: floresj@agt.cie.uma.es}}

\begin{document}
\parindent=5mm
\date{}
\maketitle

\begin{quote}

\noindent {\small \bf Abstract.}

{\small We present a new development of the causal boundary of
spacetimes, originally introduced by Geroch, Kronheimer and Penrose.
Given a strongly causal spacetime (or, more generally, a
chronological set), we reconsider the GKP ideas to construct a
family of completions with a chronology and topology extending the
original ones. Many of these completions present undesirable
features, like those which appeared in previous approaches by other
authors. However, we show that all these deficiencies are due to the
attachment of an ``excessively big'' boundary. In fact, a notion of
``completion with minimal boundary'' is then introduced in our
family such that, when we restrict to these minimal completions,
which always exist, all previous objections disappear. The optimal
character of our construction is illustrated by a number of
satisfactory properties and examples.}
\end{quote}
\begin{quote}
\end{quote}

\newpage

\section{Introduction}

Many properties in Mathematics are usually best understood by
attaching an ideal boundary to the target space. This situation is
also common in General Relativity, where important physical
questions about spacetimes are closely related with properties of
their boundaries. In the construction of such boundaries, the
causal structure of the spacetime plays a decisive role.

The most common method to place an ideal boundary on a spacetime
is by embedding it conformally into a larger spacetime and, then,
by taking the boundary of the image. The {\it conformal boundary}
was firstly introduced for asymptotically simple spacetimes in
\cite{P} and, since then, it has provided a number of interesting
insights in specific examples. However, this approach presents
several important handicaps. The construction imposes strong
mathematical restrictions on the spacetime, even though such
restrictions are satisfied by many spacetimes of physical
interest. On the other hand, there is no a systematic and totally
general way to carry out the embedding such that the standard
character of the conformal boundary is ensured.

An alternative, but similar, construction based on the new concept
of {\em isocausality} has been introduced recently in \cite{GS}. The
more accurate character of the isocausality with respect to the
causal structure of the spacetime allows the authors to generalize
the conformal approach. As a consequence, this new construction is
applicable to larger classes of spacetimes. However, it is unclear
if this method will overcome the remaining problems of the conformal
method.

The existence of a systematic and intrinsic procedure to construct
an ideal boundary for general spacetimes was first envisioned by
Geroch, Kronheimer and Penrose in \cite{GKP}. They suggested a
construction totally based on the causal structure of the spacetime.
In particular, it is invariant under conformal changes. Roughly
speaking, they placed a future (past) ideal endpoint for every
inextensible future (past) timelike curve, in such a way that it
only depends on the curve's past (future). Then, they characterized
these ideal endpoints by means of {\em terminal indecomposable past
(future) sets} TIPs (TIFs) (see Section \ref{preliminaries} for
definitions).

The GKP approach, also called {\it causal completion} or $c$-{\it
completion}, overcomes the handicaps of the conformal construction.
In fact, this method can be applied successfully to any strongly
causal spacetime (see, however, \cite{S1}), and yields a systematic
procedure to construct an unique boundary. However, this method
presents an important technical difficulty: in \cite{GKP} the
authors remarked that some TIPs and TIFs must act as the same ideal
endpoint. In particular, this makes it necessary to define {\em
non-trivial} identifications between these sets.

There are a number of papers written in order to solve this
question, which is closely related to the introduction of a
satisfactory topology for the causal completion (see \cite{GS2} for
a detailed review on the subject). The story just begins in
\cite{GKP}. The authors introduced a generalized Alexandrov topology
on the initial construction, and then, they suggested the minimum
set of identifications necessary to obtain a Hausdorff ($T_{2}$)
space, i.e. any two points can be separated by neighborhoods.
However, this method fails to produce the ``expected'' completion in
some examples \cite{KLL}, \cite{KL1}, \cite[Section 5]{MR}. On the
other hand, although strong separation properties as $T_{2}$ are
desirable, there are no physical reasons to impose it a priori. More
annoying, from a topological viewpoint the causal boundary of
Minkowski space does not match all the common expectation of a cone
\cite{H2}.

Afterwards, other more accurate attempts have been suggested. The
procedure proposed in \cite{R}, very close to the GKP approach, also
fails in simple examples (see \cite{KL2}). Another approach proposed
in \cite{BS}, and improved later in \cite{S1,S2} via the {\it
Szabados relation} (Definition \ref{Szab}; see also Section
\ref{comparing}), again presents undesirable properties (see
\cite{KL1}, \cite{KL2}, \cite[Sections 2.2, 5]{MR}).

Recently, Marolf and Ross have introduced in \cite{MR} an entirely
new use of the Szabados relation, including a new topology for the
completion, which overcomes important difficulties in previous
attempts (see Section \ref{comparing}). In fact, the MR construction
satisfies essential requirements to be considered a reasonable
completion: (i) the original spacetime becomes densely,
chronologically and topologically embedded into the completion and,
(ii) any timelike curve in the spacetime has some limit in the
completion. However, apart from certain ``anomalous'' limit
behaviors in some examples (see \cite{MR}), they also admit an
annoying failing: not only is their topology not necessarily
Hausdorff (which cannot be regarded as unsatisfactory, as we will
see later), but it might not even be $T_{1}$, i.e. some point might
not be closed (Example \ref{5}). On the other hand, the MR
completion sometimes includes too many ideal points (Example
\ref{yo}).

Another viewpoint in the study of causal completions was previously
inaugurated by Harris in \cite{H1}. Prevented by the necessity of
non-trivial identifications when considering the past and future
completions simultaneously, he only focused on the {\it future
chronological completion} $\hat{X}$ (the same study also works for
the past). In \cite{H1}, he showed the universality of this partial
completion. In \cite{H2}, he introduced a topology for $\hat{X}$
based on a limit operator $\hat{L}$, the so called {\em future
chronological topology} (see Section \ref{preliminaries}). Then, a
series of satisfactory properties for this topology were shown,
including universality when the boundary is spacelike. The specific
utility of this approach is checked in some wide classes of
spacetimes, as static and multiwarped spacetimes (see \cite{H3},
\cite{H4}, \cite{FH}; see also \cite{H5} for a review). However, the
lack of causal information by considering only a partial boundary
also implies anomalous limit behaviors in simple cases (Example
\ref{3}).

\vspace{3mm}

In this article we present a whole revision of the causal boundary
of spacetimes by combining in a totally new way the GKP ideas. Our
approach is very general, and, indeed, it includes not one, but a
family of different completions. By imposing a minimality condition,
we will choose between them those completions which are optimal,
showing that these minimal completions overcome all the deficiencies
which appeared in previous constructions by other authors.

We have included in Section \ref{preliminaries} some basic concepts
and preliminary results useful for the next sections. In order to
gain in generality, our paper does not treat just with spacetimes,
but also {\em chronological sets} $(X,\ll)$, Definition
\ref{chronrel}, a mathematical object which abstracts the
chronological structure of the spacetime.

Our approach essentially begins in Section \ref{chron}. In
Definition \ref{eend} we extend the usual notion of endpoint of a
timelike curve in a spacetime, to that of {\em endpoint} of an
arbitrary chain (totally chronologically related sequence) in a
chronological set. Then, we use this definition to introduce a
general notion of {\em completion} $\overline{X}$ for a
chronological set $X$, by imposing that any chain in $X$ admits
some endpoint in $\overline{X}$, Definition \ref{completion}.
According to this definition, now a chronological set may admit
{\em many different} completions, including the GKP {\em
pre-completion} and the {\em Marolf-Ross construction} as
particular cases.

In order to go further, our completions need also to be made
chronological sets. This is done in Section \ref{chronol}, where any
completion $\overline{X}$ is endowed with a chronological relation
$\overline{\ll}$ such that the natural inclusion ${\bf i}$ from
$(X,\ll)$ to $(\overline{X},\overline{\ll})$ becomes a {\em dense
chronological embedding}, Theorem \ref{tt8}.

With this structure of a chronological set, we can verify the
consistency of our notion of completion. This is checked in Section
\ref{joser} by showing that every completion is indeed a {\em
complete} chronological set, Definition \ref{def-compl} and Theorem
\ref{comp}.

In order to get a deeper analysis of our construction, in Section
\ref{topology} we have endowed any chronological set with a
topology, the so-called {\em chronological topology}, Definitions
\ref{overline}, \ref{closed}, which is inspired by the ideas in
\cite{H2}: first, we have introduced a (sort of theoretic-set) {\em
limit operator} $L$ on $X$, and then, defined the {\em closed sets}
as those subsets $C\subset X$ such that $L(\sigma)\subset C$ for any
sequence $\sigma\subset C$. In particular, every completion now
becomes a topological space. Then, a number of very satisfactory
properties are shown. With this topology, every chronological set
$X$ becomes {\em topologically embedded} into $\overline{X}$ via the
natural inclusion ${\bf i}$, Theorem \ref{artu}. Therefore, as the
manifold topology of a strongly causal spacetime $V$ {\em coincides}
with its chronological topology, Theorem \ref{princ}, the manifold
topology {\em is just} the restriction to $V$ of the chronological
topology of $\overline{V}$. Moreover, the notion of endpoint
previously introduced becomes now {\em compatible} (even though, non
necessarily equivalent) with the notion of {\em limit} of a chain
provided by this topology, Theorem \ref{compatible}. As a
consequence, $X$ is {\em topologically dense} in $\overline{X}$, and
any timelike curve in $V$ {\em has some limit} in $\overline{V}$,
Corollary \ref{thchar}.

All these properties show that these constructions verify essential
requirements to be considered reasonable completions. However, many
of these completions are still non-optimal, in the sense that they
have boundaries formed by ``too many'' ideal points: clearly, this
is the case of the GKP pre-completion in \cite{GKP}, which sometimes
attaches two ideal points where we would expect only one. The
existence of these spurious ideal points implicitly leads to other
undesirable features: non-equivalence between the notions of
endpoint and limit of a chain; non-closed boundaries; completions
with bad separation properties...

In order to overcome these deficiencies, in Section
\ref{universality} we have restricted our attention to those
completions $\overline{X}$ with ``the smallest boundary'': that is,
those completions which are {\em minimal} for a certain order
relation based on the ``size'' of the boundary, Definition \ref{gh}.
These minimal completions, which {\em always} exist, Theorem
\ref{ghh}, and, indeed, may be {\em non-unique} in certain cases
(Example \ref{yo}), are called {\em chronological completions},
Definition \ref{ghhh}. In Theorem \ref{7.3} these completions are
characterized in terms of some nice properties (indeed some of them
were axiomatically imposed in previous approaches). The rest of
Section \ref{universality} is devoted to show the very satisfactory
properties of these completions for strongly causal spacetimes $V$:
the notion of endpoint is now totally {\em equivalent} to that of
the limit of a chain, Theorem \ref{compatible'}; the chronological
boundaries are {\em closed} in the completions, Theorem
\ref{boundaryclosed}; the chronological completions $\overline{V}$
are always $T_{1}$, Theorem \ref{6.4}; and, even though
non-Hausdorffness is possible, violation of $T_{2}$ is exclusively
restricted to points at the boundary, Theorem \ref{non-hausdorff}.

Section \ref{comparing} has been devoted to emphasize the optimal
character of our approach by comparing it with some previous
approaches, see Theorem \ref{optm} and discussion below.

In Section \ref{applications} we have shown the utility of our
approach in Causality Theory by characterizing two levels of the
causal ladder in terms of the chronological completion: {\em
global hyperbolicity}, Theorem \ref{globhyp}; and {\em causal
simplicity}, Theorem \ref{causallysimple}.

Finally, in Section \ref{examples} we have described some simple
examples illustrating the results and properties stated in previous
sections. We have also checked our construction in two important
classes of physical spacetimes, {\em standard static spacetimes} and
{\em plane wave solutions}.

Before concluding this introduction, we remark that our approach
does not include considerations about causal, but non-chronological,
relations. We have omitted them because the essential part of the
causal structure is exclusively carried out by the chronology of the
spacetime. Thus, only chronological relations have been considered
here, and so, we have gained in simplicity. Nevertheless, the
inclusion of purely causal relations into our framework may be the
subject of future investigation.

\section{Preliminaries}\label{preliminaries}

Our construction is intended to be exclusively based on the
chronological structure of the spacetime. In order to stress this
idea, throughout this paper we will work on the simplest
mathematical object carrying this structure; the so called {\em
chronological set} (first introduced in \cite{H1}).
\begin{definition}\label{chronrel} A {\em chronological set} is a set $X$ with a
relation $\ll$ (called {\em chronological relation}) such that
$\ll$ is transitive and non-reflexive ($x\not\ll x$), there are no
isolates (everything is related chronologically to something), and
$X$ has a countable set $D$ which is dense: if $x\ll y$ then for
some $d\in D$, $x\ll d\ll y$.
\end{definition}

When working on a chronological set $X$, the role of future timelike
curves in a spacetime is now played by {\em future chains}: sequence
$\varsigma=\{x_{n}\}\subset X$ obeying $x_{1}\ll \cdots\ll x_{n}\ll
x_{n+1}\ll\cdots$. As in spacetimes, a subset $P\subset X$ is called
a {\em past} set if it coincides with its past, that is,
$P=I^{-}[P]:=\{x\in X: x\ll x'\;\hbox{for some}\; x'\in
P\}$\footnote{Here we are following standard notation: that is,
$I^{-}[\cdot]$ denotes the past of a set of points, while
$I^{-}(\cdot)$ is reserved for past of a point.}. Given a subset
$S\subset X$, we define the {\em common past} of $S$ as $\downarrow
S:=I^{-}[\{x\in X:\;\; x\ll x'\;\;\forall x'\in S\}]$. A past set
that cannot be written as the union of two proper subsets both of
which are also past sets is called an {\em indecomposable past} set
IP. (Here, the emptyset $\emptyset$ will be assumed to be a past set
which is not indecomposable.) An IP which does not coincide with the
past of any point in $X$ is called a {\em terminal indecomposable
past set} TIP. Otherwise, it is called a {\em proper indecomposable
past set} PIP. In a spacetime the past of a point is always
indecomposable, however it is easy to give examples of chronological
sets where this does not happen (Example \ref{yo}). See Figure 1 for
an illustration of these definitions.

\begin{figure}[ht]
\begin{center}
\includegraphics[width=6cm]{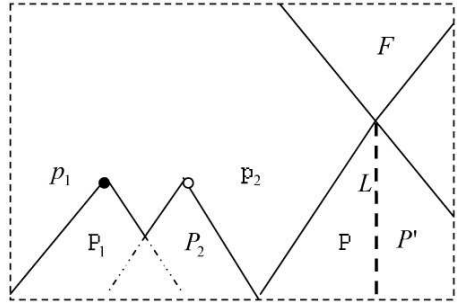}
\caption{We consider the interior region of a square in Minkowski
plane with point $p_{2}$ and segment $L$ removed: $P_{1}\cup P_{2}$
is a past set which is not indecomposable; $P_{1}, P_{2}$ are both
IPs; $P_{1}$ is a PIP ($P_{1}=I^{-}(p_{1})$) and $P_{2}$ is a TIP
($p_{2}\not\in V$); the common past $\downarrow F$ coincides with
$P\cup P'$.}
\end{center}
\end{figure}

The following adaptation of \cite[Prop. 4.1]{H2} shows that any
past set admits an useful decomposition in terms of IPs:
\begin{proposition}\label{maximality} Let $X$ be a chronological set. Every past set $\emptyset\neq P\subset X$ can be written as
$P=\cup_{\alpha}P_{\alpha}$, where $\{P_{\alpha}\}_{\alpha}$ is
the set of all maximal (under the inclusion relation) IPs included
in P.
\end{proposition}
{\it Proof.} Consider an arbitrary point $x\in P\neq\emptyset$. Let
${\cal A}_{x}$ be the set of all IPs included in $P$ which contain
$x$, endowed with the partial order of inclusion. Since $P$ is a
past set, we can construct inductively a future chain $c$ starting
at $x$ and entirely contained in $P$. In particular, the past of $c$
is an IP in ${\cal A}_{x}\neq\emptyset$. On the other hand, consider
$\{P_{i}\}_{i\in I}\subset {\cal A}_{x}$, $I$ a well-ordered index
set with $P_{i}\subset P_{k}$ for $i\leq k$. Then $\cup_{i}P_{i}$ is
clearly an IP into $P$ which also contains $x$. Whence,
$\cup_{i}P_{i}$ is an upper bound in ${\cal A}_{x}$ for
$\{P_{i}\}_{i\in I}$. Therefore, Zorn's Lemma\footnote{Zorn's Lemma:
Every non-empty partially ordered set in which every totally ordered
subset has an upper bound contains at least one maximal element.}
ensures the existence of a maximal IP $P_{x}$ into $P$ which
contains $x$. \cvd

\vspace{1mm}

\noindent Proposition \ref{maximality} justifies now the following
definition:
\begin{definition}\label{decomposition} Given a past set $\emptyset\neq P$ in a chronological set $X$, the set ${\rm
dec}(P):=\{P_{\alpha}\}_{\alpha}$ of all maximal IPs included in $P$
is called the {\em decomposition} of $P$. By convention, we will
assume $dec(\emptyset)=\emptyset$.
\end{definition}

Denote by $\hat{X}$ the set of all IPs of $X$. If $X$ is {\em
past-regular} (i.e. $I^{-}(x)$ is IP for all $x\in X$) and {\it
past-distinguishing} (i.e. $I^{-}(x)=I^{-}(x')$ implies $x=x'$),
then $\hat{X}$ is called the {\em future chronological completion}
of $X$. In this case, $\hat{X}$ can be endowed with a structure of
chronological set and the map $x\mapsto I^{-}(x)$ injects $X$ into
$\hat{X}$. Therefore, we can write $\hat{X}=X\cup
\hat{\partial}(X)$, $\hat{\partial}(X)$ being the set of all TIPs of
$X$, which is called the {\em future chronological boundary} of $X$.
More details about the future chronological completion can be found
in \cite{H1}.

\vspace{2mm}

It is possible to endow a chronological set with a topology. The
heart of the {\em future chronological topology}, firstly
introduced in \cite{H2}, is the following limit operator
$\hat{L}$:
\begin{definition}\label{hhat}
Given a sequence $\sigma=\{P_{n}\}$ of past sets in $X$, an IP
$P\subset X$ satisfies $P\in\hat{L}(\sigma)$ if
\begin{itemize} \item[(i)] $P\subset {\rm
LI}(P_{n})$ and \item[(ii)] $P$ is maximal $IP$ within ${\rm
LS}(P_{n})$,
\end{itemize}
where {\em LI} and {\em LS} denote the standard inferior and
superior limits of sets:
\[
\begin{array}{l}
{\rm
LI}(P_{n})=\liminf_{n\rightarrow\infty}(P_{n})=\cup_{n=1}^{\infty}\cap_{k=n}^{\infty}P_{k}
\\ {\rm LS}(P_{n})=\limsup_{n\rightarrow\infty}(P_{n})=\cap_{n=1}^{\infty}\cup_{k=n}^{\infty}P_{k}.
\end{array}
\]
\end{definition}
The limit operator $\hat{L}$ was introduced in \cite{H2} in a
different way. However, it is not hard to show that both
definitions are equivalent (see \cite{H5}).

Then, the {\em future chronological topology} ($\;\widehat{}\;$-{\it
topology}) of $X$ is introduced by defining the {\em closed sets} as
those subsets $C\subset X$ such that $\hat{L}(\sigma)\subset C$ for
any sequence $\sigma\subset C$\footnote{When using the limit
operator $\hat{L}$, it is common to implicitly identify the past or
future of a point with the point itself.}. With this definition,
$\hat{L}(\sigma)$ is to be thought of as first-order limits of
$\sigma$. In the particular case of $X=V$ being a strongly causal
spacetime, the $\hat{L}$-limit of a sequence coincides with the
limit with respect to the manifold topology:
\begin{proposition}\label{2.3} Let $V$ be a strongly causal
spacetime. For any sequence $\sigma=\{p_{n}\}\subset V$, a point
$p$ is $\hat{L}$-limit of $\sigma$ if and only if it is the limit
of $\sigma$ with the topology of the manifold.
\end{proposition}
{\it Proof.} See \cite[Theorem 2.3]{H2}.

\vspace{1mm}

Of course, the dual notions of the concepts introduced here (past
chain, future set, $\uparrow S$, IF, TIF, PIF, $\check{X}$,
$\check{L}$...), and the corresponding results, can be defined and
proved just by interchanging the roles of past and future.

\vspace{1mm}

We finish this section with a remarkable result coming from
\cite[Prop. 5.1]{S1}. Previously, we recall the following
definition:
\begin{definition}\label{Szab} If $P$ is maximal IP into $\downarrow F$ and $F$ is maximal IF into $\uparrow P$ then we say that $P$, $F$ are {\em S-related}, $P\sim_{S} F$ (see Figure 1).
\end{definition}
\begin{proposition}\label{propS} Let $V$ be a strongly causal spacetime. The unique S-relations involving proper
indecomposable sets in $V$ are $I^{-}(p)\sim_{S} I^{+}(p)$ for all
$p\in V$.
\end{proposition}

\section{Completing Chronological Sets}\label{chron}

A central property to be satisfied by any space $\overline{X}$
intended to be a completion of $X$ is that any chain in $X$ admits
some ``limit'' in $\overline{X}$. So, a natural strategy for
completing a chronological set consists of adding to $X$ ``ideal
endpoints'' associated to every ``endless'' chain in $X$. In order
to develop this idea, we are going to restrict our attention to
{\em weakly distinguishing} chronological sets; that is, those
chronological sets satisfying that any two points with the same
past and future must coincide. Observe that this condition is not
very restrictive at all, since it is satisfied by any strongly
causal spacetime.

Denote by $X_{p}$, $X_{f}$ the sets of all past and future sets of
$X$, resp. Then, the map
\begin{equation}\label{injection}
\begin{array}{rl}
{\bf i}: & X\rightarrow X_{p}\times X_{f} \\ & x\mapsto
(I^{-}(x),I^{+}(x))
\end{array}
\end{equation}
injects $X$ into $X_{p}\times X_{f}$ in a natural way. This
injection, joined to the fact that our construction must be
exclusively based on the chronological structure of $X$, makes
natural to conceive $\overline{X}$ as verifying
\[
{\bf i}[X]\subset\overline{X}\subset X_{p}\times X_{f}.
\]
So, if we want to completely determine $\overline{X}$, we need to
establish which elements of $X_{p}\times X_{f}$ belong to the
completion, or, equivalently, which past and future sets must be
paired to form every element of $\overline{X}$. According to the
central idea suggested at the beginning of this section, this will
be done by formalizing the notion of the ``endpoint'' of a chain. To
this aim, we are not going to define a topology on $X_{p}\times
X_{f}$. Instead, we will directly deduce a reasonable notion of
``endpoint'', which will be justified a posteriori by showing that
it is compatible with the topology for $\overline{X}$ suggested in
Section \ref{topology}.

Consider a future chain $\varsigma=\{x_{n}\}\subset X$ and assume
that it ``approaches'' to some $(P,F)\in X_{p}\times X_{f}$, where
$P$ and $F$ represent the past and future (computed in $X$) of the
limit point. If a sequence $\{p_{n}\}$ converges to some point $p$
in a spacetime, then every point in the past of $p$ is eventually in
the past of $p_{n}$. Therefore, if we translate this property to our
situation, we should obtain $P\subset I^{-}[\varsigma]$. Moreover,
since $\varsigma$ is a future chain ``approaching'' to $(P,F)$, it
becomes natural to assume $\varsigma\subset P$. In particular,
$I^{-}[\varsigma]\subset P$, and thus, $P=I^{-}[\varsigma]$. On the
other hand, by transitivity with respect to $(P,F)$, we should also
expect $F\subset \uparrow P$. Of course, there is no reason to
impose $F =\uparrow P$; however, arguing by analogy to what happens
in spacetimes, the fact that $\varsigma$ is ``approaching'' to
$(P,F)$ also leads to strengthen the inclusion $F\subset \uparrow P$
by assuming that any element in ${\rm dec}(F)$ is maximal in
$\uparrow P$, or, equivalently, ${\rm dec}(F)\subset
\check{L}(\varsigma)$. Obviously, dual conditions are deduced in the
case of $\varsigma$ being a past chain approaching to $(P,F)$.
Summarizing, we suggest the following definition:
\begin{definition}\label{eend} A pair $(P,F)\in X_{p}\times X_{f}$
is {\em endpoint} of a future (resp. past) chain $\varsigma\subset
X$ if
\begin{equation}\label{c1}
P=I^{-}[\varsigma],\quad {\rm
dec}(F)\subset\check{L}(\varsigma)\quad\qquad (\hbox{resp.}\quad
{\rm dec}(P)\subset\hat{L}(\varsigma),\quad F=I^{+}[\varsigma]).
\end{equation}
We will denote by $X^{end}$ the subset of $X_{p}\times X_{f}$
formed by the union of ${\bf i}[X]$ with all the endpoints of
every chain in $X$.
\end{definition}
Now, we are in a position to formulate the notion of {\em
completion} for a chronological set:
\begin{definition}\label{completion} Let $X$ be a weakly distinguished chronological set. A set $\overline{X}$ satisfying
\[
{\bf i}[X]\subset\overline{X}\subset X^{end}(\subset X_{p}\times
X_{f})
\]
is called a {\em completion} of $X$ if any chain in $X$ has some
endpoint in $\overline{X}$. Then, the {\em boundary} of $X$ in
$\overline{X}$ is defined as $\partial(X):=\overline{X}\setminus
{\bf i}[X]$.
\end{definition}
According to this definition, a chronological set will admit
different completions (see Example \ref{ñ}). We will denote by
${\cal C}_{X}$ the set of all these completions.

A well-known example of completion covered by Definition
\ref{completion} is the GKP {\em pre-completion} of spacetimes,
first introduced in \cite{GKP}. In fact, the pre-completion
$V^{\sharp}$ of a strongly causal spacetime $V$ can be seen as
formed by adding to ${\bf i}[V]$ the endpoints
$(I^{-}[\varsigma],\emptyset)$ (resp.
$(\emptyset,I^{+}[\varsigma])$) for every endless future (resp.
past) chain $\varsigma$\footnote{Actually, $V^{\sharp}$ was
introduced in \cite{GKP} by using identifications instead of pairs:
that is, $V^{\sharp}:=\hat{V}\cup\check{V}/\sim$, where $P\sim F$
iff $P=I^{-}(p)$, $F=I^{+}(p)$, for some $p\in V$.}. An alternative
completion is formed by adding to ${\bf i}[V]$ the endpoints $(P,F)$
given by
\begin{equation}\label{kkk}
P=I^{-}[{\rm LI}(I^{-}(x_{n}))]\quad\hbox{and}\quad F=I^{+}[{\rm
LI}(I^{+}(x_{n}))]
\end{equation}
for every endless chain $\varsigma=\{x_{n}\}$. In this case, the
resulting space $V^{\flat}$, which, in general, is different from
$V^{\sharp}$, may contain pairs (P,F) with some component $P$ or $F$
not necessarily indecomposable (see Example \ref{1}). Notice also
that $X^{end}$ is another example of completion, which, indeed,
contains any other completion of the chronological set $X$.

Observe that Definition \ref{completion} is far from being accurate:
for example, it does not avoid the possibility of having completions
which remain as completions when some boundary point is removed (see
Example \ref{ñ}). Of course, we can eliminate this possibility by
hand in Definition \ref{completion}; however, this is not sufficient
for ensuring that there are no spurious ideal points at the
boundary: for example, the GKP pre-completion $V^{\sharp}$, which
does not fall under the previous possibility, sometimes attaches two
{\em different} ideal points where we would intuitively expect only
one (see Example \ref{ñ}). We will postpone to Section
\ref{universality} the introduction of a non-trivial notion of
minimal completion, the so-called {\em chronological completion}.

Even though many completions included in Definition
\ref{completion} are not optimal, they still satisfy enough
properties to justify the name of ``completions'' for these
constructions. The next three sections are devoted to analyze
these properties in depth. To this aim, the definition and
characterizations below will be very useful:
\begin{definition}\label{gener} Let $X$ be a chronological set. A pair $(P,F)\in X_{p}\times X_{f}$ is {\em generated} by a chain $\varsigma=\{x_{n}\}\subset X$ if equalities (\ref{kkk}) hold.
\end{definition}
Of course, every chain $\varsigma$ in $X$ generates an unique pair
$(P,F)\in X_{p}\times X_{f}$.
\begin{proposition}\label{carutil} Let $X$ be a chronological set and consider a chain $\varsigma=\{x_{n}\}\subset X$ and a pair $(P,F)\in X_{p}\times X_{f}$. Then, the following statements are equivalent:
\begin{itemize}
\item[(i)] $(P,F)$ is generated by $\varsigma$; \item[(ii)] there
exists a countable dense set $D\subset X$ such that
\begin{equation}\label{conan}
P\cap D={\rm LI}(I^{-}(x_{n}))\cap D\qquad\hbox{and}\qquad F\cap
D={\rm LI}(I^{+}(x_{n}))\cap D;
\end{equation}
\item[(iii)] $(P,F)$ satisfies
\[
\begin{array}{c}
\hat{L}(\varsigma)={\rm dec}(P)\qquad\hbox{and}\qquad
\check{L}(\varsigma)={\rm dec}(F).
\end{array}
\]
\end{itemize}
\end{proposition}
{\it Proof.} We will follow this scheme: first, we will prove {\it
(iii)}$\Rightarrow${\it (ii)}; then, {\it (i)}$\Rightarrow${\it
(iii)}; and, finally, {\it (ii)}$\Rightarrow${\it (i)}.

In order to prove {\it (iii)}$\Rightarrow${\it (ii)}, first
observe that $\hat{L}(\varsigma)={\rm dec}(P)$ and
$\check{L}(\varsigma)={\rm dec}(F)$ imply
\[
P\subset {\rm LI}(I^{-}(x_{n}))\quad\hbox{and}\quad F\subset {\rm
LI}(I^{+}(x_{n})).
\]
Therefore, we only need to prove that
\begin{equation}\label{already}
P\cap D\supset {\rm LI}(I^{-}(x_{n}))\cap D\qquad\hbox{and}\qquad
F\cap D\supset {\rm LI}(I^{+}(x_{n}))\cap D
\end{equation}
for some countable dense set $D\subset X$. To this aim, take any
countable dense set $D'\subset X$ and define
$$D:=D'\setminus D_{0},$$ with $D_{0}=\{d\in D':\; I^{-}(d)\subset P\;\;\hbox{but}\;\; d\not\in P\}$.
In order to prove the density of $D$, consider $y\ll y'\in X$.
Since $D'$ is dense, there exists $d\in D'$ such that $y\ll d\ll
y'$. If $d\not\in D_{0}$, necessarily $d\in D$ and we finish.
Otherwise, consider $y\ll d$ and take $d'\in D'$ such that $y\ll
d'\ll d$. Then, necessarily $d'\not\in D_{0}$ since $d'\in
I^{-}(d)\subset P$. Therefore, $d'\in D$, and thus, $D$ is dense
in $X$.

For the first inclusion in (\ref{already}), assume by contradiction
the existence of $d\in {\rm LI}(I^{-}(x_{n}))\cap D$ such that
$d\not\in P\cap D$. From the definition of $D$, necessarily
$I^{-}(d)\not\subset P$. Therefore, there exists $x\in
I^{-}(d)\setminus P$. In particular, $x\in I^{-}[{\rm
LI}(I^{-}(x_{n}))]\neq\emptyset$, and thus, Proposition
\ref{maximality} ensures the existence of a maximal IP $P_{x}$ in
$I^{-}[{\rm LI}(I^{-}(x_{n}))]$ containing $x$. Taking into account
that $\varsigma$ is a chain, $P_{x}$ is also maximal in ${\rm
LS}(I^{-}(x_{n}))$. Therefore, $P_{x}\in \hat{L}(\varsigma)$, which
contradicts the equality $\hat{L}(\varsigma)={\rm dec}(P)$. In
conclusion, the first inclusion in (\ref{already}) holds.

We can repeat the same reasoning for the future, but taking the set
$D$ instead of $D'$ as an initial countable dense set, and removing
the elements $d\in D$ such that $I^{+}(d)\subset F$ but $d\not\in
F$. Then, the resulting set, which we also denote by $D$, clearly
satisfies both inclusions in (\ref{already}).

\vspace{2mm}

In order to prove {\it (i)}$\Rightarrow${\it (iii)}, it is clear
that $P\subset {\rm LI}(I^{-}(x_{n}))$. So, assume by
contradiction that $P_{\alpha_{0}}\in {\rm dec}(P)$ is not maximal
in ${\rm LS}(I^{-}(x_{n}))={\rm LI}(I^{-}(x_{n}))$. Then, there
exists an IP $P'$ with $P_{\alpha_{0}}\varsubsetneq P'\subset {\rm
LI}(I^{-}(x_{n}))$. In particular, $P'\not\subset P$, and thus,
there exist $x,x'\in P'\setminus P$ such that $x\ll x'$. As $x'\in
{\rm LI}(I^{-}(x_{n}))$, necessarily $x\in I^{-}[{\rm
LI}(I^{-}(x_{n}))]\setminus P$, which contradicts that
$P=I^{-}[{\rm LI}(I^{-}(x_{n}))]$. Therefore, any $P_{\alpha}\in
{\rm dec}(P)$ is maximal in ${\rm LS}(I^{-}(x_{n}))$, and thus,
${\rm dec}(P)\subset \hat{L}(\varsigma)$.

To prove $\hat{L}(\varsigma)\subset {\rm dec}(P)$, assume by
contradiction the existence of an IP $P'\in\hat{L}(\varsigma)$ such
that $P'\not\in {\rm dec}(P)$. Then, necessarily $P'\not\subset P$,
since otherwise $P'$ would be maximal in $P$ (recall that
$P=I^{-}[{\rm LI}(I^{-}(x_{n}))]$), and thus, $P'\in {\rm dec}(P)$.
Reasoning as in the previous paragraph, there exist $x,x'\in
P'\setminus P$ such that $x\ll x'$. In particular, $x'\in {\rm
LI}(I^{-}(x_{n}))$. Therefore, $x\in I^{-}[{\rm
LI}(I^{-}(x_{n}))]\setminus P$, which contradicts that $P=I^{-}[{\rm
LI}(I^{-}(x_{n}))]$. In conclusion, $\hat{L}(\varsigma)={\rm
dec}(P)$. Finally, an analogous reasoning proves that
$\check{L}(\varsigma)={\rm dec}(F)$.

\vspace{2mm}

In order to prove {\it (ii)}$\Rightarrow${\it (i)}, assume by
contradiction that $P\not\subset I^{-}[{\rm LI}(I^{-}(x_{n}))]$.
Since $P$ is a past set, necessarily $P\not\subset {\rm
LI}(I^{-}(x_{n}))$. Therefore, there exist $x,x'\in P\setminus
{\rm LI}(I^{-}(x_{n}))$ and $d\in D$ such that $x\ll d\ll x'$. In
particular, $d\in P$ but $d\not\in {\rm LI}(I^{-}(x_{n}))$.
Whence, $d$ contradicts the first equality in (\ref{conan}), and
thus, $P\subset I^{-}[{\rm LI}(I^{-}(x_{n}))]$. For the other
inclusion, assume that $x\in I^{-}[{\rm LI}(I^{-}(x_{n}))]$. This
means $x\ll x'$ for certain $x'\in {\rm LI}(I^{-}(x_{n}))$.
Therefore, there exists $d\in D$ with $x\ll d\ll x'$. In
particular, $d\in {\rm LI}(I^{-}(x_{n}))\cap D$, which, joined to
the first equality in (\ref{conan}), implies $x\ll d\in P$.
Whence, $I^{-}[{\rm LI}(I^{-}(x_{n}))]\subset P$, and the equality
follows. Finally, an analogous reasoning also shows $F=I^{+}[{\rm
LI}(I^{+}(x_{n}))]$. \cvd

\vspace{2mm}

In the next Sections \ref{chronol}--\ref{topology} by
$\overline{X}$ we will understand {\em any} completion of $X$,
according to Definition \ref{completion}.

\section{The Completions as Chronological Sets}\label{chronol}

Now that we have introduced a family of completions ${\cal C}_{X}$
for any weakly distinguishing chronological set $(X,\ll)$, the
next step consists of endowing any completion $\overline{X}\in
{\cal C}_{X}$ with a structure of weakly distinguishing
chronological set, such that $X$ becomes densely and
chronologically embedded into $\overline{X}$ via the injection
{\bf i} (see (\ref{injection})). To be more precise, let us
introduce some definitions:
\begin{definition} A bijection $f:X\rightarrow X'$ between two
chronological sets $(X,\ll)$, $(X',\ll')$ is a {\em
(chronological) isomorphism} if $f$ and $f^{-1}$ preserve the
chronological relations. When $f$ is only injective but the image
$f(X)\subset X'$ endowed with $\ll'$ is still isomorph to
$(X,\ll)$ via $f$, we say that $f$ is a {\em (chronological)
embedding} of $(X,\ll)$ into $(X',\ll')$.
\end{definition}
Consider the relation
\[
(P,F)\overline{\ll} (P',F')\quad\hbox{iff}\quad F\cap
P'\neq\emptyset,\qquad\quad \forall (P,F),(P',F')\in \overline{X}
\]
(first introduced in \cite{S1}, and used later in \cite{MR}). Then,
the following results hold:
\begin{theorem}\label{tt8} If $(X,\ll)$ is a weakly distinguishing chronological set then $(\overline{X},\overline{\ll})$ is also a chronological set. Moreover, ${\bf i}$ chronologically embeds
$(X,\ll)$ into $(\overline{X},\overline{\ll})$ in such a way that
${\bf i}[X]$ is dense in $\overline{X}$.
\end{theorem}
{\it Proof.} To prove transitivity, assume
$(P,F)\overline{\ll}(P',F')$ and $(P',F')\overline{\ll}(P'',F'')$.
Then, there exist $x\in F\cap P'$ and $x'\in F'\cap P''$. Let
$\varsigma=\{x_{n}\}\subset X$ be a chain with endpoint $(P',F')$
(if $(P',F')={\bf i}(x_{0})$ for some $x_{0}\in X$, take instead
$\varsigma=\{x_{n}\}\equiv \{x_{0}\}\subset X$). Then, $P'\subset
{\rm LI}(I^{-}(x_{n}))$, $F'\subset {\rm LI}(I^{+}(x_{n}))$. In
particular, for all $n$ big enough $x\ll x_{n}\ll x'$. But $x\in
F$ and $x'\in P''$. Hence, $x_{n}\in F\cap P''\neq\emptyset$ for
all $n$ big enough, and thus, $(P,F)\overline{\ll} (P'',F'')$.

To show that $\overline{\ll}$ is non-reflexive, assume by
contradiction that $(P,F)\overline{\ll} (P,F)$. Then, there exists
$x\in F\cap P\neq \emptyset$. As before, let
$\varsigma=\{x_{n}\}\subset X$ be a chain with endpoint $(P,F)$
(again, if $(P,F)={\bf i}(x_{0})$ for some $x_{0}\in X$, take
instead $\varsigma=\{x_{n}\}\equiv \{x_{0}\}\subset X$). Then,
$P\subset {\rm LI}(I^{-}(x_{n}))$ and $F\subset {\rm
LI}(I^{+}(x_{n}))$. In particular, for all $n$ big enough
$x_{n}\ll x\ll x_{n}$. This contradicts that $\ll$ is
non-reflexive.

To prove that there are no isolates, consider $(P,F)\in
\overline{X}$. Assume for example that $x\in P\neq\emptyset$ (if
$x\in F\neq\emptyset$, the argument is analogous). As $P$ is a
past set, there exists $x'\in P$ such that $x\ll x'$. Then, ${\bf
i}(x)\in\overline{X}$ satisfies ${\bf i}(x)\overline{\ll} (P,F)$,
since $x'\in I^{+}(x)\cap P\neq\emptyset$.

The set $${\bf i}[D]=\{{\bf i}(d): d\in D,\; D\;\hbox{countable
dense set of}\; (X,\ll)\}\subset \overline{X}$$ is a countable
dense set of $(\overline{X},\overline{\ll})$. In fact, if
$(P,F)\overline{\ll}(P',F')$ then $F\cap P'\neq\emptyset$.
Therefore, there exist $x,x'\in F\cap P'$ with $x\ll x'$. Let
$d\in D$ be such that $x\ll d\ll x'$. Then,
$(P,F)\overline{\ll}{\bf i}(d)\overline{\ll} (P',F')$, since $x\in
F\cap I^{-}(d)\neq\emptyset$ and $x'\in I^{+}(d)\cap
P'\neq\emptyset$. In particular, this also shows that ${\bf i}[X]$
is dense in $\overline{X}$.

Finally, we show that $\overline{\ll}$ extends $\ll$ without
introducing new chronological relations in ${\bf i}[X]$. Assume
first that $x,x'\in X$ satisfy $x\ll x'$. Then, there exists $d\in
D$ such that $x\ll d\ll x'$. Therefore, $d\in I^{+}(x)\cap
I^{-}(x')\neq \emptyset$ and, thus, ${\bf i}(x)\overline{\ll} {\bf
i}(x')$. Assume now that ${\bf i}(x)\overline{\ll} {\bf i}(x')$.
Then, there exists $y\in I^{+}(x)\cap I^{-}(x')\neq\emptyset$.
Therefore, $x\ll y\ll x'$, and thus, $x\ll x'$. \cvd

\begin{theorem}\label{tech} Let $(X,\ll)$ be a weakly distinguishing chronological
set. Then,\footnote{Of course, symbols $I^{-}(\cdot)$,
$I^{+}(\cdot)$ in (\ref{expected}) refer to the chronological
relation $\overline{\ll}$ instead of $\ll$.}
\begin{equation}\label{expected}
{\bf i}^{-1}[I^{-}((P,F))\cap {\bf i}[X]]=P,\quad {\bf
i}^{-1}[I^{+}((P,F))\cap {\bf i}[X]]=F\qquad\forall\,
(P,F)\in\overline{X}.
\end{equation}
In particular, $(\overline{X},\overline{\ll})$ is a weakly
distinguishing chronological set.
\end{theorem}
{\it Proof.} For the first equality in (\ref{expected}) consider
$x\in {\bf i}^{-1}[I^{-}((P,F))\cap {\bf i}[X]]$. This means that
${\bf i}(x)\overline{\ll}(P,F)\in\overline{X}$, $x\in X$. Therefore,
there exists $x'\in I^{+}(x)\cap P\neq \emptyset$, and thus, $x\ll
x'\in P$. In particular, $x\in P$. Conversely, consider $x\in P$.
Since $P$ is a past set, there exists $x'\in X$ with $x'\in
I^{+}(x)\cap P\neq\emptyset$. Therefore, ${\bf i}(x)\overline{\ll}
(P,F)$, and thus, $x\in {\bf i}^{-1}[I^{-}((P,F))\cap {\bf i}[X]]$.
The second equality in (\ref{expected}) is proved analogously.

In order to prove that $(\overline{X},\overline{\ll})$ is weakly
distinguishing, assume $I^{-}((P,F))=I^{-}((P',F'))$ and
$I^{+}((P,F))=I^{+}((P',F'))$ for some $(P,F),
(P',F')\in\overline{X}$. From (\ref{expected})
\[
\begin{array}{l}
P={\bf i}^{-1}[I^{-}((P,F))\cap {\bf i}[X]]={\bf i}^{-1}[I^{-}((P',F'))\cap {\bf i}[X]]=P' \\
F={\bf i}^{-1}[I^{+}((P,F))\cap {\bf i}[X]]={\bf
i}^{-1}[I^{+}((P',F'))\cap {\bf i}[X]]=F'.
\end{array}
\]
Whence, $(P,F)=(P',F')$. \cvd

\section{The ``Complete Character'' of the Completions}\label{joser}

In previous section we have showed that given any weakly
distinguishing chronological set $(X,\ll)$, the pair
$(\overline{X},\overline{\ll})$ is also a weakly distinguishing
chronological set. So, we can be tempted to repeat the process on
$(\overline{X},\overline{\ll})$, and construct a new pair
$(\overline{\overline{X}},\overline{\overline{\ll}})$ with
$\overline{\overline{\ll}}$ defined by
\[
({\cal P},{\cal F})\overline{\overline{\ll}}({\cal P'},{\cal
F'})\Longleftrightarrow {\cal F}\cap {\cal P'}\neq
\emptyset,\qquad\quad\forall\; ({\cal P},{\cal F}), ({\cal
P'},{\cal F'})\in \overline{\overline{X}}.
\]
In this section we are going to justify that completing a
completion is unnecessary, in the sense that any completion is
already a ``complete'' chronological set. To this aim, of course
we previously need to introduce a reasonable notion of {\em
complete} chronological set:
\begin{definition}\label{def-compl} A weakly distinguishing chronological set $(Y,\ll)$
is {\em (chronologically) complete} if ${\bf i}[Y]$ itself is a
completion of $Y$, that is, if any chain in $Y$ has some endpoint
in ${\bf i}[Y]$.
\end{definition}

\vspace{1mm}

Next, we are going to establish a suitable correspondence between
the pairs in $X_{p}\times X_{f}$ and those in
$(\overline{X})_{p}\times (\overline{X})_{f}$, for any completion
$\overline{X}$ of $X$. Consider the map
\[
\begin{array}{rl}
{\bf j}: & X_{p}\times X_{f}\rightarrow (\overline{X})_{p}\times
(\overline{X})_{f} \\ & (P,F)\mapsto (j(P),j(F)),
\end{array}
\]
where
\[
j(P):=I^{-}[{\bf i}[P]] \qquad\hbox{and}\qquad j(F):=I^{+}[{\bf
i}[F]]
\]
(here, $I^{\pm}[\cdot]$ are computed in
$(\overline{X},\overline{\ll})$). From (\ref{expected}) and the
density of ${\bf i}[X]$ into $\overline{X}$, it follows
\begin{equation}\label{nueva}
(j(P),j(F))=(I^{-}((P,F)),I^{+}((P,F)))\quad\;\;\forall\; (P,F)\in
\overline{X}.
\end{equation}
Therefore, the map ${\bf j}$ restricted to $\overline{X}\subset
X_{p}\times X_{f}$ coincides with the injection ${\bf i}$ for the
chronological set $Y=\overline{X}$. Notice also that the inverse
map of {\bf j} is given by:
\[
\begin{array}{rl}
{\bf k}: & (\overline{X})_{p}\times (\overline{X})_{f}\rightarrow
X_{p}\times X_{f} \\ & ({\cal P},{\cal F})\mapsto (k({\cal
P}),k({\cal F})),
\end{array}
\]
where
\[
k({\cal P}):={\bf i}^{-1}[{\cal P}\cap {\bf i}[X]]
\qquad\hbox{and}\qquad k({\cal F}):={\bf i}^{-1}[{\cal F}\cap {\bf
i}[X]].
\]
In fact, first notice that ${\bf k}$ is well-defined. For example,
to check that $k({\cal P})$ is a past set, consider $x\in k({\cal
P})$. This means ${\bf i}(x)\in {\cal P}$. Since ${\cal P}$ is a
past set of $\overline{X}$ and ${\bf i}[X]$ is dense in
$\overline{X}$, there exists $x'\in X$ with ${\bf
i}(x)\overline{\ll}{\bf i}(x')\in {\cal P}$. In particular, $x'\in
{\bf i}^{-1}[{\cal P}\cap {\bf i}[X]]$. Therefore, $x\ll x'\in
k({\cal P})$, showing that $k({\cal P})\subset I^{-}[k({\cal P})]$.
Conversely, assume now that $x\in I^{-}[k({\cal P})]$. Then, $x\ll
x'\in k({\cal P})$, which implies ${\bf i}(x)\ll {\bf i}(x')\in
{\cal P}$. As ${\cal P}$ is a past set, it follows ${\bf i}(x)\in
{\cal P}$, and thus, $x\in k({\cal P})$. Therefore, $I^{-}[k({\cal
P})]\subset k({\cal P})$. It rests to show that ${\bf j}$ and ${\bf
k}$ satisfy the identities
\begin{equation}\label{otronumero}
{\bf k}\circ {\bf j}=Id_{X_{p}\times X_{f}}, \qquad\qquad {\bf
j}\circ {\bf k}=Id_{(\overline{X})_{p}\times(\overline{X})_{f}}.
\end{equation}
The first identity is clearly equivalent to the equalities:
\begin{equation}\label{*1}
P=k(j(P)),\qquad F=k(j(F))\qquad\forall\; (P,F)\in X_{p}\times
X_{f}.
\end{equation}
To prove the first equality in (\ref{*1}), recall that $P$ is a
past set and ${\bf i}$ a chronological embedding. Therefore, $x\in
P$ iff ${\bf i}(x)\in {\bf i}[P]\subset I^{-}[{\bf i}[P]]=j(P)$.
But, ${\bf i}(x)\in {\bf i}[X]$. Whence, $x\in P$ iff ${\bf
i}(x)\in j(P)\cap {\bf i}[X]$. In conclusion, $x\in P$ iff $x\in
{\bf i}^{-1}[j(P)\cap {\bf i}[X]]=k(j(P))$. The second equality in
(\ref{*1}) can be proved analogously. On the other hand, the
second identity in (\ref{otronumero}) is equivalent to these other
equalities:
\begin{equation}\label{*2}
{\cal P}=j(k({\cal P})),\qquad {\cal F}=j(k({\cal
F}))\qquad\forall\; ({\cal P},{\cal F})\in
(\overline{X})_{p}\times (\overline{X})_{f}.
\end{equation}
To prove the first equality, recall that ${\cal P}$ is a past set
and ${\bf i}[X]$ is dense in $\overline{X}$. Therefore, $(P,F)\in
{\cal P}$ iff there exists $x\in X$ with $(P,F)\overline{\ll}{\bf
i}(x)\in {\cal P}$. In particular, $(P,F)\in {\cal P}$ iff
$(P,F)\overline{\ll}{\bf i}(x)\in {\bf i}[k({\cal P})]$.
Therefore, $(P,F)\in {\cal P}$ iff $(P,F)\in j(k({\cal P}))$. The
second equality in (\ref{*2}) can be proved analogously.

\vspace{1mm}

With these tools, we are now ready to prove Theorem \ref{comp}
below. The hard part of this proof has been extracted in the
following lemma:
\begin{lemma}\label{completeness} Let $\overline{X}$ be a completion of a weakly distinguishing chronological set $(X,\ll)$.
\begin{itemize}
\item[(i)] If $(P,F)\in X_{p}\times X_{f}$ is an endpoint of a chain
$\delta=\{x_{i}\}\subset X$, then the pair $(j(P),j(F))\in
(\overline{X})_{p}\times (\overline{X})_{f}$ is an endpoint of the
chain ${\bf i}[\delta]=\{{\bf i}(x_{i})\}_{i}\subset \overline{X}$.
\item[(ii)] Given a chain $\varsigma\subset\overline{X}$ there
exists another chain $\delta\subset X$ such that ${\bf
i}[\delta]\subset\overline{X}$ and $\varsigma$ have the same
endpoints.
\end{itemize}
\end{lemma}
{\it Proof.} For {\it (i)}, we assume without restriction that
$(P,F)$ is an endpoint of a future chain $\delta=\{x_{i}\}\subset
X$. From Definitions \ref{eend}, \ref{gener}, this implies
\begin{equation}\label{5.0}
P=P'\quad\hbox{and}\quad {\rm dec}(F)\subset {\rm dec}(F'),
\end{equation}
for $(P',F')$ being the pair generated by $\delta$. From
Proposition \ref{carutil} there exists a countable dense set
$D\subset X$ such that
\begin{equation}\label{5.4}
\begin{array}{l}
P'\cap D={\rm LI}(I^{-}(x_{i}))\cap D \\ F'\cap D={\rm
LI}(I^{+}(x_{i}))\cap D.
\end{array}
\end{equation}
Taking into account that ${\bf i}:X\hookrightarrow\overline{X}$ is
a dense and chronological embedding, from (\ref{5.0}) and
(\ref{5.4}) we easily obtain
\begin{equation}\label{k1}
j(P)=j(P')\quad\hbox{and}\quad {\rm dec}(j(F))\subset {\rm
dec}(j(F')),
\end{equation}
with
\begin{equation}\label{k2}
\begin{array}{l}
j(P')\cap {\bf i}[D]={\rm LI}(I^{-}({\bf i}(x_{i})))\cap
{\bf i}[D] \\
j(F')\cap {\bf i}[D]={\rm LI}(I^{+}({\bf i}(x_{i})))\cap {\bf
i}[D].
\end{array}
\end{equation}
From (\ref{k2}) and Proposition \ref{carutil}, we deduce that
$(j(P'),j(F'))$ is generated by ${\bf
i}[\delta]\subset\overline{X}$. This joined to (\ref{k1}) proves
that $(j(P),j(F))$ is an endpoint of ${\bf i}[\delta]$ (recall again
Definitions \ref{eend}, \ref{gener}).

\vspace{2mm}

In order to prove {\it (ii)}, we can assume without restriction
that $\varsigma=\{(P_{n},F_{n})\}_{n}\subset\partial(X)$ is a
future chain. Let $\varsigma^{o}=\{(P_{n}^{o},F_{n}^{o})\}\subset
X^{end}$ be a future chain formed by pairs $(P_{n}^{o},F_{n}^{o})$
generated by some chain $\varsigma^{n}=\{x^{n}_{m}\}_{m}\subset X$
admitting some endpoint equal to $(P_{n},F_{n})$. In particular,
\begin{equation}\label{true}
P_{n}\subset P_{n}^{o}\quad\hbox{and}\quad F_{n}\subset
F_{n}^{o}\quad \hbox{for all}\;\; n.
\end{equation}
Denote by $({\cal P}^{o},{\cal F}^{o})\in (X^{end})_{p}\times
(X^{end})_{f}$ the pair generated by $\varsigma^{o}$. First we are
going to prove the existence of some future chain
$\delta=\{x_{i}\}\subset X$ generating $(k({\cal P}^{o}),k({\cal
F}^{o}))$.

From Proposition \ref{carutil} and (\ref{nueva}) we have
\begin{equation}\label{005}
\begin{array}{l}
{\cal P}^{o}\cap {\bf i}[D]={\rm
LI}(I^{-}((P^{o}_{n},F^{o}_{n})))\cap {\bf i}[D]={\rm
LI}(j(P^{o}_{n}))\cap {\bf i}[D]
\\ {\cal F}^{o}\cap {\bf i}[D]={\rm LI}(I^{+}((P^{o}_{n},F^{o}_{n})))\cap {\bf i}[D]={\rm LI}(j(F^{o}_{n}))\cap {\bf
i}[D].
\end{array}
\end{equation}
Applying ${\bf i}^{-1}$ to (\ref{005}), from (\ref{*1}) we deduce
\begin{equation}\label{007}
\begin{array}{l}
k({\cal P}^{o})\cap D={\rm LI}(k(j(P^{o}_{n})))\cap D={\rm
LI}(P^{o}_{n})\cap D
\\ k({\cal F}^{o})\cap D={\rm LI}(k(j(F^{o}_{n})))\cap D={\rm LI}(F^{o}_{n})\cap D.
\end{array}
\end{equation}
Therefore, if $\varsigma^{o}\subset {\bf i}[X]$ then $\delta={\bf
i}^{-1}[\varsigma^{o}]$ is the required sequence. Otherwise,
observe that chains $\varsigma^{n}=\{x^{n}_{m}\}_{m}\subset X$
satisfy
\begin{equation}\label{gota}
P^{o}_{n}=I^{-}[{\rm LI}(I^{-}(x^{n}_{m}))]\;\;\; \hbox{and}\;\;\;
F^{o}_{n}=I^{+}[{\rm LI}(I^{+}(x^{n}_{m}))]
\end{equation}
(recall Definition \ref{gener}). In order to construct the
announced chain $\delta=\{x_{i}\}_{i}\subset X$, we will argue
inductively:

\vspace{2mm}

{\em Step 1.} Consider $d_{1}\in D$. If $d_{1}\in k({\cal P}^{o})$
(resp. $d_{1}\in k({\cal F}^{o})$), from (\ref{007}) we can define
a sequence $\{n^{1}_{k}\}_{k}\subset \N$ by removing from
$\{n\}_{n}$ those elements $n$ with $d_{1}\not\in P^{o}_{n}$
(resp. $d_{1}\not\in F^{o}_{n}$). Moreover, from (\ref{gota}) we
can construct a sequence $\{m^{1}_{k,l}\}_{l}\subset \N$ by
removing from $\{m\}_{m}$ those elements $m$ with $d_{1}\not\ll
x^{n^{1}_{k}}_{m}$ (resp. $x^{n^{1}_{k}}_{m}\not\ll d_{1}$). With
these definitions, $d_{1}\in I^{-}(x^{n^{1}_{k}}_{m^{1}_{k,l}})$
(resp. $d_{1}\in I^{+}(x^{n^{1}_{k}}_{m^{1}_{k,l}})$) for all
$k,l$. If $d_{1}\not\in k({\cal P}^{o})\cup k({\cal F}^{o})$
define $\{n^{1}_{k}\}_{k}\equiv \{n\}_{n}$,
$\{m^{1}_{k,l}\}_{l}\equiv \{m\}_{m}$.

\vspace{2mm}

{\it Step 2:} Assume now that $\{n^{i}_{k}\}_{k},
\{m^{i}_{k,l}\}_{l}\subset\N$ have been defined for certain $i$.
Consider $d_{i+1}\in D$. If $d_{i+1}\in k({\cal P}^{o})$ (resp.
$d_{i+1}\in k({\cal F}^{o})$), from (\ref{007}) we can define a
sequence $\{n^{i+1}_{k}\}_{k}\subset \N$ by removing from
$\{n^{i}_{k}\}_{k}$ those elements $n^{i}_{k}$ with
$d_{i+1}\not\in P^{o}_{n^{i}_{k}}$ (resp. $d_{i+1}\not\in
F^{o}_{n^{i}_{k}}$). Moreover, from (\ref{gota}) we can construct
a sequence $\{m^{i+1}_{k,l}\}_{l}\subset \N$ by removing from
$\{m^{i}_{k,l}\}_{l}$ those elements $m^{i}_{k,l}$ with
$d_{i+1}\not\ll x^{n^{i+1}_{k}}_{m^{i}_{k,l}}$ (resp.
$x^{n^{i+1}_{k}}_{m^{i}_{k,l}}\not\ll d_{i+1}$). With these
definitions, $d_{i+1}\in I^{-}(x^{n^{i+1}_{k}}_{m^{i+1}_{k,l}})$
(resp. $d_{i+1}\in I^{+}(x^{n^{i+1}_{k}}_{m^{i+1}_{k,l}})$) for
all $k,l$. If $d_{i+1}\not\in k({\cal P}^{o})\cup k({\cal F}^{o})$
define $\{n^{i+1}_{k}\}_{k}\equiv \{n^{i}_{k}\}_{k}$,
$\{m^{i+1}_{k,l}\}_{l}\equiv \{m^{i}_{k,l}\}_{l}$. Therefore, we
can construct by induction $\{n^{i}_{k}\}_{k}$,
$\{m^{i}_{k,l}\}_{l}$ for all $i\in\N$. Moreover, it is possible
to choose $l(i)$ in such a way that $\delta=\{x_{i}\}_{i}\equiv
\{x^{n^{i}_{i}}_{m^{i}_{i,l(i)}}\}_{i}$ is a future chain
contained in $k({\cal P}^{o})$.

\vspace{1mm}

With this definition of $\delta$, the following inclusions hold:
\begin{equation}\label{assumption2}
\begin{array}{c}
{\rm LI}(P^{o}_{n})\cap D\subset {\rm LI}(I^{-}(x_{i}))\cap
D\subset {\rm LI}(P^{o}_{n})\cap D
\\ {\rm LI}(F^{o}_{n})\cap D\subset
{\rm LI}(I^{+}(x_{i}))\cap D\subset {\rm LI}(F^{o}_{n})\cap D.
\end{array}
\end{equation}
In fact, assume $d\in {\rm LI}(P^{o}_{n})\cap D$. From (\ref{007})
and previous construction, necessarily $d\ll
x^{n^{i}_{k}}_{m^{i}_{k,l}}$ for all $i$ big enough and all $k,l$.
In particular, $d\ll x_{i}$ for all $i$ big enough, and thus,
$d\in {\rm LI}(I^{-}(x_{i}))\cap D$. This proves the first
inclusion in (\ref{assumption2}). Assume now $d\in {\rm
LI}(I^{-}(x_{i}))\cap D$. As $\delta\subset k({\cal P}^{o})$,
necessarily $d\in k({\cal P}^{o})$. Therefore, from (\ref{007}) we
have $d\in {\rm LI}(P^{o}_{n})\cap D$. For the inclusions in the
second line of (\ref{assumption2}), assume first $d\in {\rm
LI}(F^{o}_{n})\cap D$. Reasoning as before, we deduce $d\gg
x^{n^{i}_{k}}_{m^{i}_{k,l}}$ for all $i$ big enough and all $k,l$,
and thus, $d\in {\rm LI}(I^{+}(x_{i}))\cap D$. Assume now $d\in
{\rm LI}(I^{+}(x_{i}))\cap D$. There exists $d_{i_{n}}\in k({\cal
P}^{o})\cap F^{o}_{n}\neq\emptyset$ for all $n$. But,
$d_{i_{n}}\ll x_{i}$ for all $i\geq i_{n}$. Whence, $d\gg
d_{i_{n}}\in F^{o}_{n}$ for all $n$. This proves the last
inclusion in (\ref{assumption2}).

In conclusion, from (\ref{007}) and (\ref{assumption2}) we have
\begin{equation}\label{isstar}
\begin{array}{c}
k({\cal P}^{o})\cap D={\rm LI}(I^{-}(x_{i}))\cap D \\
k({\cal F}^{o})\cap D={\rm LI}(I^{+}(x_{i}))\cap D.
\end{array}
\end{equation}
Therefore, from Proposition \ref{carutil} we conclude that
$(k({\cal P}^{o}),k({\cal F}^{o}))$ is generated by $\delta$.

Next, it remains to show that ${\bf i}[\delta]\subset\overline{X}$
and $\varsigma$ have the same endpoints. Taking into account that
${\bf i}$ is a chronological embedding, from (\ref{*2}) and
(\ref{isstar}) we deduce
\[
\begin{array}{l}
{\cal P}^{o}\cap {\bf i}[D]=jk({\cal P}^{o})\cap {\bf i}[D]={\rm
LI}(I^{-}({\bf i}(x_{i})))\cap {\bf i}[D] \\ {\cal F}^{o}\cap {\bf
i}[D]=jk({\cal F}^{o})\cap {\bf i}[D]={\rm LI}(I^{+}({\bf
i}(x_{i})))\cap {\bf i}[D].
\end{array}
\]
Whence, Proposition \ref{carutil} ensures that $({\cal
P}^{o},{\cal F}^{o})$ is generated by ${\bf i}[\delta]$. Since
$({\cal P}^{o},{\cal F}^{o})$ is also generated by
$\varsigma^{o}$, for every $i_{0}$, $n_{0}$ there exists $n$, $i$
big enough such that
\begin{equation}\label{112}
{\bf i}(x_{i_{0}})\overline{\ll}
(P^{o}_{n-1},F^{o}_{n-1})\qquad\hbox{and}\qquad
(P^{o}_{n_{0}+1},F^{o}_{n_{0}+1})\overline{\ll} {\bf i}(x_{i}).
\end{equation}
Moreover, from (\ref{true}) and the fact that $\varsigma$ is a
future chain, necessarily
\begin{equation}\label{111}
(P^{o}_{n-1},F^{o}_{n-1})\overline{\ll}
(P_{n},F_{n})\qquad\hbox{and}\qquad
(P_{n_{0}},F_{n_{0}})\overline{\ll}
(P^{o}_{n_{0}+1},F^{o}_{n_{0}+1}).
\end{equation}
Therefore, taking into account (\ref{112}) and (\ref{111}) we have
proved that fixed for $i_{0}$, $n_{0}$ there exists $n$, $i$ big
enough such that
\[
{\bf i}(x_{i_{0}})\overline{\ll}
(P_{n},F_{n})\qquad\hbox{and}\qquad
(P_{n_{0}},F_{n_{0}})\overline{\ll} {\bf i}(x_{i}).
\]
In conclusion, $\varsigma$ and ${\bf i}[\delta]$ have the same
endpoints. \cvd

\vspace{2mm}

Now, the main result of this section can be proved easily:
\begin{theorem}\label{comp} (Completeness). If $(X,\ll)$ is a weakly distinguishing chronological set
then $(\overline{X},\overline{\ll})$ is complete.
\end{theorem}
{\it Proof.} Given any chain
$\varsigma=\{(P_{n},F_{n})\}\subset\overline{X}$ we need to prove
that $\varsigma$ has some endpoint in ${\bf j}[\overline{X}]$. From
Lemma \ref{completeness} {\it (ii)} there exists some chain
$\delta\subset X$ such that ${\bf i}[\delta]\subset \overline{X}$
and $\varsigma$ have the same endpoints. Let $(P,F)$ be some
endpoint of $\delta$ in $\overline{X}$. From Lemma
\ref{completeness} {\it (i)}, $(j(P),j(F))$ is an endpoint of ${\bf
i}[\delta]$. Therefore, $(j(P),j(F))\in {\bf j}[\overline{X}]$ is
also endpoint of $\varsigma$, as required. \cvd

\section{The Chronological Topology}\label{topology}

In order to provide a deeper description of the relation between a
chronological set and its boundary, in this section we are going
to introduce a topological structure. More precisely, we are going
to endow any chronological set with the {\em chronological
topology}, a non-trivial generalization of the
$\;\widehat{}\;$-topology in \cite{H2}.

To this aim, first we are going to define a limit operator $L$ for
any chronological set $Y$. We take as a guide property the fact that
our topology must turn the endpoints of chains (Definition
\ref{eend}) into topological limits. So, the following definition,
based on a simple generalization of conditions (\ref{c1}), becomes
natural:
\begin{definition}\label{overline}  Given a sequence
$\sigma\subset Y$, we say that $x\in L(\sigma)$ if
\[
{\rm dec}(I^{-}(x))\subset \hat{L}(\sigma)\quad\hbox{and}\quad {\rm
dec}(I^{+}(x))\subset \check{L}(\sigma).
\]
\end{definition}
With this limit operator in hand we can now define the {\em
closed} sets of $Y$, which determine the {\em chronological
topology} ($chr$-{\it topology}):
\begin{definition}\label{closed} The {\em closed sets} of $Y$ with the $chr$-{\em topology} are those subsets $C\subset Y$ such that $L(\sigma)\subset C$ for any sequence $\sigma\subset C$.
\end{definition}

\vspace{1mm}

It is worth noting that the $chr$-topology has been defined for {\em
any} chronological set. In particular, it is applicable to both
strongly causal spacetimes $V$ and their completions $\overline{V}$.
So, two natural questions arise: does the manifold topology of a
strongly causal spacetime coincide with the $chr$-topology it
inherits when considered as a chronological set?; does the manifold
topology of a strongly causal spacetime $V$ coincide with the
restriction to $V$ of the $chr$-topology of $\overline{V}$? The
following two theorems will answer positively to these questions.
For the second one, the hypothesis of strong causality (further than
weakly distinguishing) becomes essential.

\begin{theorem}\label{artu} Any weakly-distinguishing chronological set $(X,\ll)$ is topologically embedded
into $(\overline{X},\overline{\ll})$ via the injection {\bf i} if
both spaces are endowed with the $chr$-topology.
\end{theorem}
{\it Proof.} It suffices to show that a point $x\in X$ satisfies
$x\in L(\sigma)$ for some sequence $\sigma=\{x_{n}\}\subset X$ if
and only if ${\bf i}(x)\in L(\rho)$ for $\rho=\{{\bf
i}(x_{n})\}\subset\overline{X}$. Taking into account that ${\bf
i}:X\hookrightarrow\overline{X}$ is a dense and chronological
embedding, and equalities (\ref{*1}), (\ref{*2}), we deduce:
\[
\begin{array}{l}
P\in {\rm dec}(I^{-}(x)),\;\; P\subset {\rm LI}(I^{-}(x_{n}))
\\ P\;\;\hbox{maximal in}\;\; {\rm
LS}(I^{-}(x_{n}))
\end{array}
\Rightarrow
\begin{array}{l}
j(P)\in {\rm dec}(I^{-}({\bf i}(x))),\;\; j(P)\subset {\rm
LI}(I^{-}({\bf i}(x_{n})))
\\ j(P)\;\;\hbox{maximal in}\;\; {\rm
LS}(I^{-}({\bf i}(x_{n}))),
\end{array}
\]
\[
\begin{array}{l}
{\cal P}\in {\rm dec}(I^{-}({\bf i}(x))),\;\; {\cal P}\subset {\rm
LI}(I^{-}({\bf i}(x_{n})))
\\ {\cal P}\;\;\hbox{maximal in}\;\; {\rm
LS}(I^{-}({\bf i}(x_{n})))
\end{array}
\Rightarrow
\begin{array}{l}
k({\cal P})\in {\rm dec}(I^{-}(x)),\;\; k({\cal P})\subset {\rm
LI}(I^{-}(x_{n}))
\\ k({\cal P})\;\;\hbox{maximal in}\;\; {\rm
LS}(I^{-}(x_{n})).
\end{array}
\]
Analogously, we deduce the corresponding implications for the
future. Therefore, the thesis follows from (\ref{nueva}) and
Definitions \ref{hhat}, \ref{overline}. \cvd

\begin{theorem}\label{princ} The topology of a strongly causal spacetime $V$ as a manifold coincides with the corresponding $chr$-topology.
\end{theorem}
{\it Proof.} From Proposition \ref{2.3} (and its dual), a point
$p\in V$ is the limit of a sequence $\sigma=\{p_{n}\}\subset V$
with the topology of the manifold if and only if
\begin{equation}\label{uyy}
I^{-}(p)\in\hat{L}(\sigma)\qquad\hbox{and}\qquad
I^{+}(p)\in\check{L}(\sigma).
\end{equation}
Taking into account that ${\rm dec}(I^{-}(p))=\{I^{-}(p)\}$ and
${\rm dec}(I^{+}(p))=\{I^{+}(p)\}$, conditions (\ref{uyy}) can be
written as
\[
{\rm dec}(I^{-}(p))\subset \hat{L}(\sigma)\quad\hbox{and}\quad {\rm
dec}(I^{+}(p))\subset \check{L}(\sigma).
\]
Therefore, from Definition \ref{overline}, $\sigma$ converges to $p$
with the manifold topology if and only if $p$ is the $L$-limit of
$\sigma$. \cvd

%

\vspace{2mm}

With this topology, we can also prove that the concept of {\em
endpoint} is compatible with the notion of {\em limit} of a chain:
\begin{theorem}\label{compatible} Let $\varsigma=\{x_{n}\}$ be a chain
in a weakly distinguishing chronological set $X$. Then, the
following statements hold:
\begin{itemize}
\item[(1)] If ${\bf i}(x)\in {\bf i}[X]$ is endpoint of
$\varsigma$ then $x\in L(\varsigma)$. Moreover, the reciprocal is
true if, in addition, $X$ is regular (i.e., past- and
future-regular). \item[(2)] If $(P,F)\in\overline{X}$ is an endpoint
of $\varsigma$ then $(P,F)\in L(\rho')$ for any subsequence
$\rho'\subset \rho=\{{\bf i}(x_{n})\}_{n}\subset {\bf i}[X]$. In
particular, ${\bf i}[X]$ is topologically dense in $\overline{X}$.
\end{itemize}
\end{theorem}
{\it Proof.} Statement {\it (1)} is a direct consequence of
Definitions \ref{eend}, \ref{overline}.

For {\it (2)}, assume without restriction that $\varsigma$ is a
future chain. From Definitions \ref{eend}, \ref{gener},
\begin{equation}\label{?}
P=P^{o}\quad\hbox{and}\quad {\rm dec}(F)\subset {\rm dec}(F^{o}),
\end{equation}
where $(P^{o},F^{o})$ is the pair generated by $\varsigma$. From
Proposition \ref{carutil} there exists a countable dense set
$D\subset X$ such that
\begin{equation}\label{lla}
P^{o}\cap D={\rm LI}(I^{-}(x_{n}))\cap D,\qquad F^{o}\cap D={\rm
LI}(I^{+}(x_{n}))\cap D.
\end{equation}
Taking into account that ${\bf i}$ is a chronological embedding,
from (\ref{lla}) we obtain:
\begin{equation}\label{virt}
\begin{array}{l}
j(P^{o})\cap {\bf i}[D]={\rm LI}(I^{-}({\bf i}(x_{n})))\cap {\bf
i}[D]
\\ j(F^{o})\cap {\bf i}[D]={\rm LI}(I^{+}({\bf
i}(x_{n})))\cap {\bf i}[D].
\end{array}
\end{equation}
Moreover, equalities (\ref{virt}) also hold for any subsequence
$\rho'\subset \rho=\{{\bf i}(x_{n})\}$ since $\rho$ is a chain in
$\overline{X}$. This joined to (\ref{?}) and Proposition
\ref{carutil} imply
\[
\{j(P)\}=\{j(P^{o})\}=\hat{L}(\rho') ,\quad {\rm
dec}(j(F))\subset{\rm
dec}(j(F^{o}))=\check{L}(\rho')\quad\hbox{for any}\;\;
\rho'\subset\rho,
\]
and thus, $(P,F)\in L(\rho')$ for any $\rho'\subset\rho$. \cvd

\vspace{2mm}

\noindent Finally, as a direct consequence of Theorem
\ref{compatible} {\it (1)} and Theorem \ref{comp} we obtain the
following result:
\begin{corollary}\label{thchar} If $\varsigma$ is a chain in a complete chronological set $Y$ then $\varsigma$ has some limit in $Y$. In particular, if $Y=\overline{V}$
is a completion for some strongly causal spacetime $V$, then any
timelike curve $\gamma(\equiv {\bf i}[\gamma])$ in $V(\equiv {\bf
i}[V])$ has some limit in $\overline{V}$.
\end{corollary}

\vspace{2mm}

The results obtained so far in this paper show that our construction
satisfies essential requirements to provide reasonable completions
for any strongly causal spacetime. However, many of these
completions are not optimal, in the sense that they include spurious
ideal points. As a consequence: the notions of endpoint and limit of
chains, even if compatible, are not totally equivalent (Example
\ref{ñ}), the boundaries may not be closed in the completions
(Example \ref{3}), the completions may present bad separation
properties (Example \ref{ñ})... In the next section we are going to
show that all these deficiencies disappear when only completions
with minimal boundaries are considered.

\section{The Chronological Completions}\label{universality}

In order to look for completions with minimal boundaries, first we
delete from ${\cal C}_{X}$ those completions which are still
completions when some point of its boundary is removed. Denote by
${\cal C}_{X}^{*}$ the resulting set, which is always non-empty:
for example, ${\bf i}[X]$ joined to those pairs of the form
$(I^{-}[\varsigma],\emptyset)$ or $(\emptyset,I^{+}[\varsigma])$
for any future or past chain $\varsigma$ without endpoints in
${\bf i}[X]$, is a completion in ${\cal C}_{X}^{*}$. Then, we
introduce a partial order relation $\leq$ in ${\cal C}_{X}^{*}$.
Roughly speaking, we will say that $\overline{X}^{\imath}$ {\em
precedes} $\overline{X}^{\jmath}$ if there exists a suitable
partition of $\partial^{\jmath}(X)$ by $\partial^{\imath}(X)$.
More precisely:
\begin{definition}\label{gh} Let $X$ be a weakly
distinguishing chronological set and consider two completions
$\overline{X}^{\imath}, \overline{X}^{\jmath}\in {\cal C}_{X}^{*}$
with
\[
\partial^{\imath}(X)=\{(P_{i},F_{i}): i\in
I\}\qquad\hbox{and}\qquad
\partial^{\jmath}(X)=\{(P_{j},F_{j}): j\in J\}.
\]
Then, we say $\overline{X}^{\imath}\leq\overline{X}^{\jmath}$ if
there exists some partition $\partial^{\jmath}(X)=\cup_{i\in
I}S_{i}$, $S_{i}\cap S_{i'}=\emptyset$ if $i\neq i'$, satisfying
the following conditions:
\begin{itemize}
\item[(i)] if a chain in $X$ has some endpoint in $S_{i}$ then
$(P_{i},F_{i})$ is also an endpoint of that chain,\footnote{In
particular, observe that every $S_{i}$ has at most two elements.}
and \item[(ii)] for every $i\in I$ such that $S_{i}=\{(P,F)\}$ with
$(P,F)$, $(P_{i},F_{i})$ endpoints of the same chains, it is ${\rm
dec}(P_{i})\subset {\rm dec}(P)$ and ${\rm dec}(F_{i})\subset {\rm
dec}(F)$.
\end{itemize}
\end{definition}
With this definition the pair $({\cal C}_{X}^{*},\leq)$ becomes a
partially ordered set ({\em reflexivity} and {\em transitivity}
are direct; {\em antisymmetry} needs a simple discussion involving
several cases). Furthermore, $({\cal C}_{X}^{*},\leq)$ always
admits some minimal element:
\begin{theorem}\label{ghh} If $X$ is a weakly distinguishing chronological set then $({\cal C}_{X}^{*},\leq)$ has some minimal element.
\end{theorem}
{\it Proof.} We can assume without restriction that any completion
$\overline{X}\in {\cal C}_{X}^{*}$ satisfies that any pair $(P,F)\in
\partial(X)$ has ${\rm dec}(P)$ and ${\rm dec}(F)$ finite
(otherwise, remove from ${\cal C}_{X}^{*}$ those completions which
do not verify this property; the minimal elements of the resulting
set, which is non-empty, are still minimal elements of ${\cal
C}_{X}^{*}$).

Consider $\{\overline{X}^{\imath_{\alpha}}\}_{\alpha\in
\Lambda}\subset {\cal C}_{X}^{*}$, $\Lambda$ a well-ordered set with
$\overline{X}^{\imath_{\alpha}}\leq\overline{X}^{\imath_{\beta}}$
for $\alpha\geq\beta$. Fix any $\alpha_{0}\in\Lambda$, let ${\cal
S}$ be a set of chains in $X$ such that
\[
\partial^{\imath_{\alpha_{0}}}(X)=\{(P^{\varsigma}_{\alpha_{0}},F^{\varsigma}_{\alpha_{0}}): \varsigma\in {\cal
S}\},
\]
with $(P_{\alpha_{0}}^{\varsigma},F_{\alpha_{0}}^{\varsigma})$
being some endpoint of $\varsigma$ in
$\overline{X}^{\imath_{\alpha_{0}}}$. Then, ${\cal S}$ also
satisfies
\[
\partial^{\imath_{\alpha}}(X)=\{(P^{\varsigma}_{\alpha},F^{\varsigma}_{\alpha}): \varsigma\in {\cal
S}\},\qquad\hbox{for all}\;\; \alpha\geq\alpha_{0},
\]
with $(P^{\varsigma}_{\alpha},F^{\varsigma}_{\alpha})$ being the
pre-image in $\partial^{\imath_{\alpha}}(X)$ of
$(P^{\varsigma}_{\alpha_{0}},F^{\varsigma}_{\alpha_{0}})$ via some
partition of $\partial^{\imath_{\alpha_{0}}}(X)$ by
$\partial^{\imath_{\alpha}}(X)$ according to Definition \ref{gh}.
With these definitions,
$\{P^{\varsigma}_{\alpha}\}_{\alpha\geq\alpha_{0}}$,
$\{F^{\varsigma}_{\alpha}\}_{\alpha\geq\alpha_{0}}$ are necessarily
constants for all $\alpha\geq\alpha^{*}$, for some
$\alpha^{*}\in\Lambda$ depending on $\varsigma$. Therefore, the set
\[
\overline{X}:={\bf i}[X]\cup
\{(P^{\varsigma}_{\alpha^{*}},F^{\varsigma}_{\alpha^{*}}):
\varsigma\in {\cal S}\}\in {\cal C}_{X}^{*}
\]
is a lower bound for
$\{\overline{X}^{\imath_{\alpha}}\}_{\alpha\in \Lambda}$, and
thus, Zorn's Lemma ensures the existence of some minimal
completion in ${\cal C}_{X}^{*}$. \cvd

\vspace{1mm}

We are now ready to introduce the notion of {\em chronological
completion}:
\begin{definition}\label{ghhh} A completion $\overline{X}$ in ${\cal C}_{X}$ is a {\em chronological completion} if it is a minimal element of $({\cal
C}_{X}^{*},\leq)$. Then, the chronological boundary of $X$ in
$\overline{X}$ is defined as $\partial(X):=\overline{X}\setminus
{\bf i}[X]$.
\end{definition}
Even if it is not very common, there are spacetimes admitting
different chronological completions (see Example \ref{yo}).
However, if $X$ is complete then ${\cal C}_{X}^{*}=\{{\bf
i}[X]\}$, and thus, ${\bf i}[X]$ is the unique chronological
completion of $X$.

The next result establishes a series of nice properties (some of
them axiomatically imposed in previous approaches) which totally
characterize these constructions:
\begin{theorem}\label{7.3} Let $V$ be a strongly causal spacetime. Then, a subset $\partial(V)\subset V_{p}\times V_{f}$ is the chronological boundary associated to some chronological completion $\overline{V}$ of $V$ if and only if the following properties hold:
\begin{itemize}
\item[(1)] Every terminal indecomposable set in $V$ is
the component of some pair in $\partial(V)$. Moreover, if
$(P,F)\in\partial(V)$ then $P$, $F$ are both indecomposable sets if
non-empty.
\item[(2)] If $P,F\neq\emptyset$ satisfy $(P,F)\in\partial(V)$
then $P\sim_{S} F$. \item[(3)] If $(P,\emptyset)\in\partial(V)$
(resp. $(\emptyset,F)\in\partial(V)$) then $P$ (resp. $F$) is not
S-related to anything. \item[(4)] If $(P,F)\in\partial(V)$ then $P$,
$F$ are both terminal sets if non-empty. \item[(5)] If $(P,F_{1}),
(P,F_{2})\in\partial(V)$, $F_{1}\neq F_{2}$ (resp. $(P_{1},F)$,
$(P_{2},F)\in\partial(V)$, $P_{1}\neq P_{2}$) then $F_{i}$ (resp.
$P_{i}$), $i=1,2$, do not appear in any other pair of $\partial(V)$.
\end{itemize}
\end{theorem}
{\it Proof.} First, we will prove the implication to the right.

{\em (1)} If, for example, some TIP $P\neq\emptyset$ is not the
component of any pair in $\partial(V)$, then any future chain
$\varsigma\subset V$ with $I^{-}[\varsigma]=P$ has no endpoint in
$\overline{V}$, which contradicts that $\overline{V}$ is a
completion.

For the second assertion, assume that $(P,F)\in\partial(V)$ is an
endpoint of some future chain $\varsigma\subset V$. From Definition
\ref{eend}, it is $P=I^{-}[\varsigma]$, and thus, $P$ is IP. Assume
by contradiction that $F\neq\emptyset$ is not IF. Then, if we
replace the pair $(P,F)$ in $\overline{V}$ by $(P,\emptyset)$, the
resulting set is still a completion which contradicts the minimal
character of $\overline{V}$.

{\em (2)} By contradiction, assume for example that
$(P,F)\in\partial(V)$ is an endpoint of some future chain, but $P$
is not maximal IP into $\downarrow F$. Since $\overline{V}$ is a
completion, there exists some past set $P'\neq P$ such that
$(P',F)\in\overline{V}$. Therefore, if we replace $(P,F)$ in
$\overline{V}$ by $(P,\emptyset)$, the resulting set is still a
completion which contradicts the minimal character of
$\overline{V}$.

{\em (3)} By contradiction, assume that
$(P,\emptyset)\in\partial(V)$ but $P\sim_{S} F$ for some IF $F$.
Then, if we replace $(P,\emptyset)$ in $\overline{V}$ by $(P,F)$,
the resulting set is still a completion which contradicts the
minimal character of $\overline{V}$.

{\em (4)} It directly follows from {\em (2)}, {\em (3)} and
Proposition \ref{propS}.

{\em (5)} Assume by contradiction that $(P,F_{1}), (P,F_{2}),
(P',F_{1})\in\partial(V)$, with $F_{1}\neq F_{2}$ and $P\neq P'$.
Then, if we remove the pair $(P,F_{1})$ from $\overline{V}$, the
resulting set is still a completion, and thus, contradicts that
$\overline{V}\in {\cal C}_{V}^{*}$.

\vspace{1mm}

Conversely, consider $\overline{V}:={\bf i}[V]\cup \partial(V)$ with
$\partial(V)$ satisfying conditions {\it (1)}--{\it (5)}. From {\it
(1)} and {\it (2)}, $\overline{V}$ is a completion. From {\it (1)},
{\it (2)}, {\it (3)} and {\it (5)}, it is $\overline{V}\in {\cal
C}_{V}^{*}$. In order to prove that $\overline{V}$ is minimal in
$({\cal C}_{V}^{*},\leq)$, assume that
$\overline{V}^{\imath}\leq\overline{V}$ for some completion
$\overline{V}^{\imath}\in {\cal C}_{V}^{*}$. Then, there exists some
partition $\partial(V)=\cup_{i\in I}S_{i}$, $S_{i}\cap
S_{i'}=\emptyset$ if $i\neq i'$, satisfying condition {\it (i)},
{\it (ii)} in Definition \ref{gh}. From {\it (1)}, {\it (2)} and
{\it (3)}, if $(P,F)\in S_{i}$ then $P=P_{i}$, $F=F_{i}$. Whence,
$S_{i}=\{(P_{i},F_{i})\}$ for all $i\in I$, and thus,
$\overline{V}^{\imath}=\overline{V}$. \cvd

\vspace{2mm}

When $X=V$ is a strongly causal spacetime, the chronological
completions verify a number of very satisfactory properties. We
begin by showing the equivalence between the notions of {\em
endpoint} and {\em limit} of a chain:
\begin{theorem}\label{compatible'} Let $\varsigma=\{p_{n}\}$ be a chain
in a strongly causal spacetime $V$. A pair $(P,F)$ is an endpoint of
$\varsigma$ in some chronological completion $\overline{V}$ if and
only if $(P,F)\in L(\rho')$ for any subsequence $\rho'\subset
\rho=\{{\bf i}(p_{n})\}_{n}\subset {\bf i}[V]$.
\end{theorem}
{\it Proof.} From Theorem \ref{compatible} {\it (2)}, we only need
to prove the implication to the left. So, assume $(P,F)\in L(\rho)$
for $\rho=\{{\bf i}(p_{n})\}_{n}\subset {\bf i}[V]$. If
$P\neq\emptyset$ (resp. $F\neq\emptyset$) and $\varsigma$ is a
future (resp. past) chain then $P=I^{-}[\varsigma]$ (resp.
$F=I^{+}[\varsigma]$), and thus, the implication directly follows
from Definitions \ref{eend}, \ref{overline}. So, assume for example
that $(\emptyset,F)\in\partial(V)$ is a $L$-limit of the future
chain $\rho$. From Definition \ref{overline} and Theorem \ref{7.3}
{\it (1)}, $F$ is maximal IF into $\uparrow I^{-}[\varsigma]$. Let
$P'$ be a maximal IP into $\downarrow F$ containing
$I^{-}[\varsigma]$. Then, $P'\sim_{S} F$, and thus, Theorem
\ref{7.3} {\it (3)} implies $(\emptyset,F)\not\in
\partial(V)$, a contradiction. Whence, this last possibility cannot
happen. \cvd

%

\vspace{2mm}

Moreover, the chronological boundary is always closed in the
corresponding chronological completion:
\begin{theorem}\label{boundaryclosed} If $V$ is a strongly causal spacetime then $\partial(V)$ is closed in
$\overline{V}$.
\end{theorem}
{\it Proof.} By contradiction, assume the existence of a sequence
$\sigma=\{(P_{n},F_{n})\}\subset \partial(V)$ such that ${\bf
i}(p)\in L(\sigma)$ for some $p\in V$. For every $n$, consider a
chain $\varsigma^{n}=\{p^{n}_{m}\}_{m}\subset V$ with endpoint
$(P_{n},F_{n})$. Then,
\begin{equation}\label{peli}
\hbox{either}\quad
\hat{L}(\varsigma^{n})=\{I^{-}[\varsigma^{n}]\}=\{P_{n}\}\quad\hbox{or}\quad
\check{L}(\varsigma^{n})=\{I^{+}[\varsigma^{n}]\}=\{F_{n}\},
\end{equation}
depending on if $\varsigma^{n}$ is either future or past chain,
resp. Let $U\subset V$ be a precompact neighborhood of $p$. For
every $n$, necessarily $\{p^{n}_{m}\}_{m}\subset
V\setminus\overline{U}$ eventually for all $m$. In fact, otherwise
$\varsigma^{n}$ converges (up to a subsequence) to certain
$r_{n}\in \overline{U}$ with the topology of the manifold, and
thus, $I^{-}(r_{n})\in \hat{L}(\varsigma^{n})$ and
$I^{+}(r_{n})\in \check{L}(\varsigma^{n})$ (Proposition \ref{2.3}
and its dual). Therefore, from (\ref{peli}), either
$P_{n}=I^{-}(r_{n})$ or $F_{n}=I^{+}(r_{n})$, in contradiction
with $(P_{n},F_{n})\in\partial(V)$ (recall Theorem \ref{7.3} {\it
(4)}). In conclusion, fixed future and past chains
$\varsigma=\{q_{k}\}$, $\varsigma'=\{q'_{k}\}$ such that
$I^{-}(p)=I^{-}[\varsigma]$ and $I^{+}(p)=I^{+}[\varsigma']$, we
can choose $\{n_{k}\}_{k}$, $\{m_{k}\}_{k}$ satisfying $q_{k}\ll
p^{n_{k}}_{m_{k}}\ll q'_{k}$ and $p^{n_{k}}_{m_{k}}\in V\setminus
\overline{U}$ for all $k$. Therefore, taking into account that
$\varsigma$, $\varsigma'$ converge to $p$ with the topology of the
manifold, any sequence of future-directed timelike curves joining
$q_{k}$ with $p^{n_{k}}_{m_{k}}$ and then with $q'_{k}$
contradicts the strong causality of $V$. \cvd

\vspace{2mm}

The chronological completions also satisfy reasonably good
separation properties. In fact, from Definition \ref{overline},
Proposition \ref{propS} and Theorem \ref{7.3} {\it (1)}, {\it (2)},
{\it (3)} every element $(P,F)\in\overline{V}$ is the unique limit
in $\overline{V}$ of the sequence constantly equal to $(P,F)$, thus:
\begin{theorem}\label{6.4} If $V$ is a strongly causal spacetime then $\overline{V}$ is $T_{1}$.
\end{theorem}

\noindent Notice however that $\overline{V}$ is not always $T_{2}$
(Example \ref{1}). The lack of Hausdorffness in chronological
completions cannot be attributed to a defect of our particular
approach. On the contrary, it seems a remarkable property
intrinsic to the causal boundary approach itself (see \cite{MR}
for an interesting discussion on this question).

Even if the chronological completions may be non-Hausdorff, there
are still some restrictions to the elements of $\overline{V}$ which
can be non-Hausdorff related, Theorem \ref{non-hausdorff}. In order
to prove this result, we need the following proposition:
\begin{proposition}\label{intermedio} Let $\overline{V}$ be a
chronological completion of a strongly causal spacetime $V$. If
$K\subset V$ is compact in $V$ then ${\bf i}[K]$ is closed in
$\overline{V}$.
\end{proposition}
{\it Proof.} From Theorems \ref{artu}, \ref{princ}, ${\bf i}[K]$ is
closed in ${\bf i}[V]$. So, by contradiction, we will assume the
existence of $(P,F)\in
\partial(V)$, with $(P,F)\in L(\rho)$ for a certain sequence
$\rho=\{{\bf i}(p_{n})\}$ such that $\sigma=\{p_{n}\}\subset K$.
Since $K$ is compact, we can also assume that
$\sigma=\{p_{n}\}\subset K$ converges to some $p\in K$.

First, observe that $P,F\neq \emptyset$. In fact, by contradiction,
assume for example that $F=\emptyset$. Let $F'$ be a maximal IF in
$\uparrow P$ containing $I^{+}(p)$. Then, necessarily $P$ is maximal
IP into $\downarrow F'$, and thus, $P\sim_{S}F'$, which contradicts
Theorem \ref{7.3} {\it (3)}. Moreover, from Theorem \ref{7.3} {\em
(2)} it is also $P\sim_{S} F$. Whence, it cannot happen that
$P\subset I^{-}(p_{n})$ and $F\subset I^{+}(p_{n})$ for some $n$;
so, assume for example that
\begin{equation}\label{u1}
P\not\subset I^{-}(p_{n})\quad\hbox{for infinitely many}\;\; n.
\end{equation}
Let $\gamma$ be a future-directed timelike curve with
$P=I^{-}[\gamma]$. Since $(P,F)\in L(\rho)$, necessarily
\begin{equation}\label{u2}
I^{-}[\gamma]=P\subset {\rm LI}(I^{-}(p_{n})).
\end{equation}
Up to a subsequence of $\sigma$, from (\ref{u1}) and (\ref{u2}) we
can choose a future chain $\varsigma=\{r_{n}\}_{n}\subset\gamma$
with $I^{-}[\varsigma]=I^{-}[\gamma]$ and
$r_{n}\in\dot{I}^{-}(p_{n})$ for all $n$ ($\dot{I}^{-}(p_{n})$
denotes the topological boundary of $I^{-}(p_{n})$) such that $q\ll
p_{n}$ for all $q\ll r_{n}$. These conditions necessarily imply
$P\varsubsetneq I^{-}(p)\subset \downarrow F$, which contradicts
that $P$ is maximal IP into $\downarrow F$. \cvd


\begin{theorem}\label{non-hausdorff} Let $\overline{V}$ be a chronological completion of a strongly causal spacetime $V$. If two elements of $\overline{V}$ are
non-Hausdorff related then they are both in $\partial(V)$.
\end{theorem}
{\it Proof.} Let $\sigma=\{(P_{n},F_{n})\}\subset\overline{V}$ be
a sequence such that $(P,F),(P',F')\in L(\sigma)$ with $(P,F)\neq
(P',F')$. Assume by contradiction that $(P,F)=
(I^{-}(p),I^{+}(p))$ for some $p\in V$. From Theorems
\ref{boundaryclosed}, \ref{artu} and \ref{princ}, it is not a
restriction to assume $(P_{n},F_{n})=(I^{-}(p_{n}),I^{+}(p_{n}))$
for all $n$, with $\{p_{n}\}_{n}\subset V$ converging to $p$ with
the topology of the manifold. In particular, $K=\{p_{n}\}_{n}\cup
\{p\}$ is a compact set in $V$. Therefore, from Proposition
\ref{intermedio}, {\bf i}[K] is closed in $\overline{V}$, and
thus, $(P',F')\in {\bf i}[K]$. Again from Theorems \ref{artu},
\ref{princ}, this contradicts that $(P,F)\neq (P',F')$. \cvd

\vspace{2mm}

We finish this section by remarking that all these satisfactory
results do not avoid the existence of some ``contra-intuitive''
limit behaviors for the $chr$-topology. Consider the situation
described in Example \ref{3}. If we appeal to our intuition,
inherited from the natural embedding of this space into Minkowski,
we would expect that the sequence $\sigma$ converges to
$(P_{0},F_{0})\in\partial(V)$, which is not the case for the
$chr$-topology. Notice however that this intuition is using
additional information not exclusively contained in the causal
structure of the spacetime. More precisely, if we analyze the
chronology of the elements of $\sigma$, we observe that all of them
have empty future. But the future in $V$ of $(P_{0},F_{0})$ is
$F_{0}\neq \emptyset$. Therefore, any topology exclusively based on
the chronology must conclude that $\sigma$ does {\em not} converge
to $(P_{0},F_{0})$, as the $chr$-topology does (notice that this
situation is totally different from that showed by the examples in
\cite[Sect. III]{KL1}, \cite[Sect. II, III]{KL2}, where the
chronology of the elements of the sequences have a good limit
behavior, but, still, there is no convergence with the topologies
involved there). This discussion shows that the causal boundary
approach should not be considered an innocent variation of the
conformal boundary. On the contrary, it provides a genuine insight
on the asymptotic causal structure of the spacetime. In this sense,
Examples \ref{1} and \ref{3} tell us that the asymptotic causal
structure of $\L^{2}$ is modified in a very different way if we
remove a vertical segment than if we remove a horizontal one (this
is reasonable since time evolves just in the vertical direction), in
contraposition with the conformal boundary approach, which does not
reflect this asymmetry. Therefore, even though this limit behavior
does not reproduce the situation in the conformal boundary, we
consider this difference very satisfactory.

\section{Comparison with Other Approaches}\label{comparing}

A natural question still needs to be investigated in order to
emphasize the optimal character of our construction: what is the
relation between the chronological completions and the completions
suggested by other authors? In order to fix ideas we have chosen
what perhaps are the most accurate approaches to the (total)
causal boundary, up to date: the {\em Marolf-Ross} and the {\em
Szabados} completions.

The Marolf-Ross completion $\overline{V}_{MR}$ of a strongly
causal spacetime $V$ is formed by all the pairs $(P,F)$ composed
by an IP $P$ and an IF $F$, such that: (i) $P\sim_{S} F$, or (ii)
$P=\emptyset$ and $P'\not\sim_{S}F$ for any IP $P'$, or (iii)
$F=\emptyset$ and $P\not\sim_{S}F'$ for any IF $F'$. The
chronology adopted here is also
\[
(P,F)\overline{\ll} (P',F')\quad\hbox{iff}\quad F\cap
P'\neq\emptyset.
\]
So, taking into account Proposition \ref{propS}, the Marolf-Ross
construction becomes the biggest completion (according to Definition
\ref{completion}) which satisfies properties {\it (1)}--{\it (4)} in
Theorem \ref{7.3}. In particular, if $V$ admits more than one
chronological completion then $\overline{V}_{MR}$ is {\em strictly
greater} than any of them, since it includes the union of all of
them, as illustrated in the first spacetime of Example \ref{yo}.
Remarkably, the strict inclusion $\overline{V}\varsubsetneq
\overline{V}_{MR}$ may also hold even when $V$ admits just one
chronological completion, as illustrated in the second spacetime of
Example \ref{yo}.

The authors also adopt a topology for $\overline{V}_{MR}$: the
topology generated by the subbasis $\overline{V}_{MR}\setminus
L^{\pm}(S)$ for any $S\subset \overline{V}_{MR}$, where
\[
\begin{array}{c}
L^{+}(S)=Cl_{FB}[S\cup L^{+}_{IF}(S)] \\ L^{-}(S)=Cl_{PB}[S\cup
L^{-}_{IP}(S)]
\end{array}
\]
with
\[
\begin{array}{c}
Cl_{FB}(S)=S\cup\{(P,F)\in\overline{V}_{MR}: F=\emptyset,
P\in\hat{L}(P_{n})\;\;\hbox{for}\;\; (P_{n},F'_{n})\in S\} \\
Cl_{PB}(S)=S\cup\{(P,F)\in\overline{V}_{MR}: P=\emptyset,
F\in\check{L}(F_{n})\;\;\hbox{for}\;\; (P'_{n},F_{n})\in S\}
\end{array}
\]
and
\[
\begin{array}{c}
L^{+}_{IF}(S)=\{(P,F)\in\overline{V}_{MR}: F\neq\emptyset,
F\subset\cup_{(P',F')\in S}F'\} \\
L^{-}_{IP}(S)=\{(P,F)\in\overline{V}_{MR}: P\neq\emptyset,
P\subset\cup_{(P',F')\in S}P'\}.
\end{array}
\]
With these structures, the following remarkable comparison result
can be stated:
\begin{theorem}\label{optm} Given a strongly causal spacetime $V$, the inclusion defines a continuous and chronological map from every chronological completion $\overline{V}$ into the MR completion $\overline{V}_{MR}$.
\end{theorem}
{\it Proof.} Let $\overline{V}$ be any chronological completion of
$V$. As $\overline{V}\subset\overline{V}_{MR}$, the inclusion
$i:\overline{V}\hookrightarrow\overline{V}_{MR}$ is well-defined
and is always chronological. In order to prove that $i$ is also
continuous, suppose that $\{(P_{n},F_{n})\}_{n}\subset
\overline{V}$ converges to some $(P,F)\in\overline{V}$ with the
$chr$-topology. We wish to prove that any open set
$U=\overline{V}_{MR}\setminus L^{\pm}(S)$,
$S\subset\overline{V}_{MR}$, of the subbasis which generates the
MR topology such that $(P,F)\in U$ necessarily contains
$(P_{n},F_{n})$ for all $n$ big enough. First, notice that
$(P_{n},F_{n})\not\in S\cup L^{+}_{IF}(S)$ for all $n$ big enough.
In fact, otherwise, $F\subset {\rm
LI}(F_{n})\subset\cup_{(P',F')\in S}F'$. So, if $F\neq\emptyset$
then $(P,F)\in L^{+}_{IF}(S)\subset L^{+}(S)$, a contradiction.
If, instead, $F=\emptyset$, taking into account that $P\in
\hat{L}(P_{n})$, necessarily $(P,F)\in Cl_{FB}[S\cup
L^{+}_{IF}(S)]=L^{+}(S)$, which is again a contradiction.
Therefore, $(P_{n},F_{n})\not\in S\cup L^{+}_{IF}(S)$ for all $n$
big enough. Furthermore, $(P_{n},F_{n})$ cannot be
$(P_{n},\emptyset)$, $P_{n}\in\hat{L}(P_{k}^{n})$,
$(P_{k}^{n},F_{k}^{n})\in S\cup L^{+}_{IF}(S)$ for all $n$ big
enough. In fact, otherwise, it must be $F=\emptyset$ and
$P\in\hat{L}(P^{n}_{k_{n}})$ for some subsequence
$\{k_{n}\}_{n}\subset \{k\}_{k}$, with
$(P^{n}_{k_{n}},F^{n}_{k_{n}})\in S\cup L^{+}_{IF}(S)$, and thus,
$(P,F)\in Cl_{FB}[S\cup L^{+}_{IF}(S)]=L^{+}(S)$, a contradiction.
In conclusion, $(P_{n},F_{n})\not\in Cl_{FB}[S\cup
L^{+}_{IF}(S)]=L^{+}(S)$ for all $n$ big enough, as required. \cvd

\vspace{2mm}

A conceptually different approach to the causal boundary of
spacetimes consists of using identifications instead of pairs to
form the ideal points of the boundary (see \cite{GKP}; and the
subsequent papers \cite{R,BS,S1,S2}). This approach presents
important objections (see for example \cite[Section 2.2]{MR} for an
interesting discussion); however, sometimes some identifications may
be useful to emphasize certain aspects of the original spacetime.
Our purpose here is to provide some evidence that chronological
completions are the optimal spaces on which to establish eventual
identifications. To this aim, we are going to give an improved
version of the Szabados construction just by establishing some
natural identifications on any chronological completion.

The Szabados completion $\overline{V}_{S}$ of a strongly causal
spacetime $V$ is formed by taking the quotient of $V^{\sharp}$ (as
defined in footnote 4) by the minimum equivalence relation ${\cal
R}$ containing $\sim_{S}$\footnote{According to \cite{S2} further
identifications between pairs of TIPs or TIFs must be considered;
however, they will be omitted in our discussion.}. Whence, each
point $m\in\overline{V}_{S}$ is a class
$[P_{1},P_{2},\ldots;F_{1},F_{2},\ldots]$ of ${\cal R}$-equivalent
IPs and IFs. Szabados writes $m\ll m'$ if, for some $F_{\alpha}\in
\pi^{-1}(m)$ and $P'_{\mu}\in\pi^{-1}(m')$, $F_{\alpha}\cap
P'_{\mu}\neq\emptyset$. He also endows $\overline{V}_{S}$ with the
quotient topology of ${\cal T}^{\sharp}$ under ${\cal R}$, where
${\cal T}^{\sharp}$ is the {\em extended Alexandrov topology}
defined on $V^{\sharp}$, that is, the coarsest topology such that
for each $A\in\check{V}$, $B\in \hat{V}$ the four sets $A^{{\rm
int}}$, $B^{{\rm ext}}$, $B^{{\rm int}}$, $A^{{\rm ext}}$ are open
sets, where
\[
\begin{array}{l}
A^{\rm int}=\{P^{*}\in V^{\sharp}: P\in \hat{V}\;\hbox{and}\; P\cap
A\neq\emptyset\}, \\ A^{\rm ext}=\{P^{*}\in V^{\sharp}: P\in
\hat{V}\;\hbox{and}\; \forall S\subset V\; P=I^{-}[S]\Rightarrow
I^{+}[S]\not\subset A \}
\end{array}
\]
(the sets $B^{{\rm int}}$ and $B^{{\rm ext}}$ have similar
definitions with the roles of past and future interchanged).

In order to compare any chronological completion $\overline{V}$ with
the Szabados completion $\overline{V}_{S}$, the following
identifications on $\overline{V}$ become natural: two pairs in
$\overline{V}$ are $R$-related iff some of its respective components
are ${\cal R}$-related. Then, we endow the resulting quotient space
$\overline{V}/R$ with the corresponding quotient structures; that
is, the quotient chronology
\[
[(P_{1},F_{1})]\overline{\ll} [(P_{2},F_{2})]\quad\hbox{iff}\quad
(P_{1}',F_{1}')\overline{\ll} (P_{2}',F_{2}')\;\;\hbox{for
some}\;\;\left\{\begin{array}{l}
(P_{1}',F_{1}')\,R\,(P_{1},F_{1}) \\
(P_{2}',F_{2}')\,R\,(P_{2},F_{2}); \end{array}\right.
\]
and the quotient of the $chr$-topology.

There is an obvious bijection
$b:\overline{V}/R\rightarrow\overline{V}_{S}$ which maps every class
$[(P,F)]\in\overline{V}/R$ to the class $m\in \overline{V}_{S}$
formed by all the IPs and IFs appearing in some pair in $[(P,F)]$.
With this definition, $b$ is obviously a chronological isomorphism.
Furthermore, the only examples where $b$ is not continuous seem to
be exclusively caused by ``pathologies'' of the Szabados topology,
and thus, cannot be regarded as an anomaly of our construction. An
illustrative example of this situation is showed in \cite[Sect.
III]{KL2}.

\section{Causal Ladder and the Boundary of Spacetimes}\label{applications}

The causal boundary approach constitutes an useful tool for
examining the causal nature of a spacetime ``at infinity'', which
usually reflects important global aspects of the causal structure.
Therefore, it may be interesting to analyze the causality of a
spacetime just by looking at the boundary.

An illustrative example of this situation is the following
characterization of global hyperbolicity: {\em a spacetime is
globally hyperbolic iff there are no elements at the boundary whose
past and future are both non-empty.} This result, proposed in
\cite{Se} in a slightly different context, was proved by Budic and
Sachs in \cite[Th. 6.2]{BS}. However, their proof lies on the
particular approach of the authors to the causal boundary (developed
also in \cite{BS}), and thus, it suffers from the same important
restriction: it is only valid for {\em causally continuous}
spacetimes.

The main aim of this section consists of extending this
characterization to any strongly causal spacetime by using the
chronological boundary of spacetimes. More precisely, we prove:
\begin{theorem}\label{globhyp} Let $V$ be a strongly causal spacetime. Then, $V$ is globally
hyperbolic if and only if there are no elements
$(P,F)\in\partial(V)$ with $P,F\neq\emptyset$.
\end{theorem}
{\it Proof.} First, recall that a strongly causal spacetime is
globally hyperbolic if and only if the causal diamond
$J(p,q):=J^{+}(p)\cap J^{-}(q)$ is compact for any $p,q\in V$. As
a direct consequence of \cite[p. 409, Lemma 14]{O}, this holds if
$J(p,q)$ is included in a compact set $K\subset V$.

For the implication to the left, assume that there are no elements
$(P,F)\in\partial(V)$ with $P,F\neq\emptyset$. Take
$K=\overline{I(p,q)}\supset J(p,q)$ with $I(p,q):=I^{+}(p)\cap
I^{-}(q)$. Therefore, in order to prove that $K$ is compact we only
need to show that any sequence in $I(p,q)$ admits a subsequence with
some limit in $V$. By contradiction, assume that some $\sigma\subset
I(p,q)$ does not satisfy this assertion. From \cite[Theorem
5.11]{FH} applied to $\hat{V}$, there exists some subsequence
$\sigma'\subset\sigma$ and some TIP $P$ such that $\emptyset\neq
I^{-}(p)\subset P$ and $P\in\hat{L}(\sigma')$. Let $\emptyset\neq F$
be a maximal TIF into $\uparrow P$. Let $\emptyset\neq P\subset P'$
be some maximal TIP into $\downarrow F$. Then, necessarily
$P'\sim_{S} F$. Therefore, from Theorem \ref{7.3} ({\it 1}), ({\it
3}), there exists some IF $F'\neq\emptyset$, such that
$(P',F')\in\partial(V)$. This contradicts the hypothesis on the
boundary. Whence, $K$ is compact.

For the implication to the right, assume the existence of
$(P,F)\in\partial(V)$ with $P,F\neq\emptyset$. Take points $p\in
P$, $q\in F$. Take a chain $\varsigma\subset V$ with endpoint
$(P,F)$. Then, the elements of $\varsigma$ are eventually
contained in $I(p,q)$. However, from Theorem \ref{compatible} {\it
(2)} and Theorem \ref{non-hausdorff}, there cannot exist
subsequences of $\varsigma$ converging in $V$. Whence,
$J(p,q)\subset V$ is not compact. \cvd

\vspace{1mm}

\noindent A simple illustration of this result is provided by the
Minkowski plane $\L^{2}$: any element $(P,F)\in \partial(\L^{2})$
satisfies either $P=\emptyset$ or $F=\emptyset$ (Example \ref{0});
however, when a point is removed, and thus, the spacetime is no
longer globally hyperbolic, a pair $(P,F)$ with $P,F\neq\emptyset$
immediately appears (Example \ref{ñ}).

\vspace{2mm}

The causal boundary approach also becomes useful to characterize
other levels of the causal ladder as {\em causal simplicity},
\cite[Cor. 5.2]{BS}. For completeness, we are going to prove an
extension of this result, again valid for any strongly causal
spacetime. To this aim, consider any causal relation $\prec$ on any
chronological completion $\overline{V}$ such that over ${\bf i}[V]$
it satisfies:\footnote{We have introduced here causal relations just
to establish Theorem \ref{causallysimple}, but with no further
pretension. We postpone to a future paper a more precise definition
of causal relations into our framework.}
\[
{\bf i}(p)\prec {\bf i}(q)\quad\hbox{iff}\quad\hbox{either}\;\;
I^{+}(q)\subset I^{+}(p)\;\;\hbox{or}\;\; I^{-}(p)\subset I^{-}(q).
\]
Then, the following results hold:
\begin{lemma}\label{las} For any two points $p,q$ in a strongly causal spacetime $V$, ${\bf i}(p)\prec {\bf i}(q)$ if and only if, either $q\in
\overline{J^{+}(p)}$ or $p\in \overline{J^{-}(q)}$.
\end{lemma}
{\it Proof.} For the implication to the right, assume that ${\bf
i}(p)\prec {\bf i}(q)$. If $I^{+}(q)\subset I^{+}(p)$, take any
sequence $\{q_{n}\}\subset I^{+}(q)$ such that $q_{n}\rightarrow q$.
Then, we obtain
$q\in\overline{I^{+}(q)}\subset\overline{I^{+}(p)}\subset\overline{J^{+}(p)}$.
If, instead, $I^{-}(p)\subset I^{-}(q)$, just reason analogously to
obtain
$p\in\overline{I^{-}(p)}\subset\overline{I^{-}(q)}\subset\overline{J^{-}(q)}$.

For the implication to the left, first assume that $q\in
\overline{J^{+}(p)}$. Take $q'\in I^{+}(q)$ and a sequence
$\{q_{n}\}\subset J^{+}(p)$ with $q_{n}\rightarrow q$. For all $n$
big enough, $q_{n}\ll q'$. Whence, $q'\in I^{+}(p)$, and thus,
$I^{+}(q)\subset I^{+}(p)$. Therefore, ${\bf i}(p)\prec {\bf
i}(q)$. Assume now that $p\in\overline{J^{-}(q)}$. Reasoning
analogously we deduce $I^{-}(p)\subset I^{-}(q)$, and thus, ${\bf
i}(p)\prec {\bf i}(q)$ holds again. \cvd

\begin{theorem}\label{causallysimple} A strongly causal spacetime $V$ is causally simple if and only if the
causality $\prec$ of $\overline{V}$ restricted to ${\bf i}[V]$
coincides with that of $V$.
\end{theorem}
{\it Proof.} For the implication to the right, Lemma \ref{las}
implies that relation ${\bf i}(p)\prec {\bf i}(q)$ holds if and only
if either $q\in\overline{J^{+}(p)}$ or $p\in\overline{J^{-}(q)}$,
and, from the hypothesis, this holds if and only if $q\in J^{+}(p)$.
Therefore, the causality of $\overline{V}$ restricted to ${\bf
i}[V]$ coincides with that of $V$.

For the implication to the left, first assume that $q\in
\overline{J^{+}(p)}$. From Lemma \ref{las}, it is ${\bf i}(p)\prec
{\bf i}(q)$, and, from the hypothesis, this implies $q\in
J^{+}(p)$. Therefore, $J^{+}(p)$ is closed. Analogously, if $p\in
\overline{J^{-}(q)}$, necessarily $p\in J^{-}(q)$, and thus,
$J^{-}(q)$ is closed too. Therefore, $V$ is causally simple. \cvd

\vspace{2mm}

Finally, let us remark that the link between the causal ladder and
the boundary of spacetimes may also arise in a deeper way. In
Example \ref{10} we have indicated how the dramatic change in the
level of causality of generalized wave type spacetimes
(\ref{genwav}), when metric coefficient $-H$ leaves the quadratic
growth, translates into a low dimensionality of the chronological
boundary for these spacetimes. It would be interesting to explore if
this intriguing relation between the critical behavior of the
causality of a spacetime and the dimensionality of its boundary is
generalizable to further classes of spacetimes.

\section{Examples}\label{examples}

In this section we briefly examine our construction in some
examples and compare it with previous approaches, putting special
emphasis in the differences between them.

\vspace{1mm}

\begin{example}\label{0} {\em Consider Minkowski space $V=\L^{n+1}(\equiv \R^{n}\times\L^{1})$.
In order to construct the chronological boundary $\partial(V)$, in
this case it suffices to consider the set of pairs
$(I^{-}[\gamma],\emptyset)$ (resp. $(\emptyset,I^{+}[\gamma])$), for
every inextensible future-directed (resp. past-directed) lightlike
geodesic $\gamma$, in addition to the pairs $(V,\emptyset)$ and
$(\emptyset,V)$. Therefore, $\partial(V)$ can be identified with a
pair of cones on $S^{n-1}$ with apexes $i^{+}$, $i^{-}$ (Figure 2).
This is in total agreement with the image of the (standard)
conformal embedding of Minkowski space into the Einstein Static
Universe.

\begin{figure}[ht]
\begin{center}
\includegraphics[width=5cm]{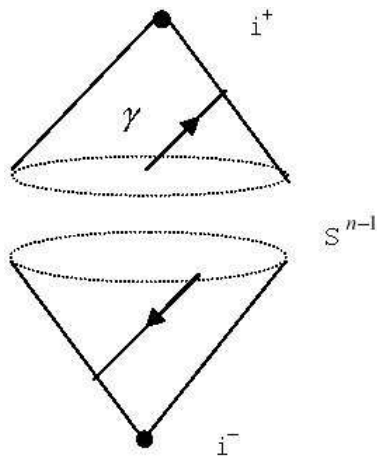}
\caption{Chronological boundary for $\L^{n+1}$.}
\end{center}
\end{figure}

On the other hand, the limit of every sequence in $\overline{V}$
with the $chr$-topology coincides with the set-theoretic limit of
the elements of the sequence. Again, this provides just the same
topology as that inherited from the (standard) conformal embedding
of $\L^{n+1}$ into ESU.}

\end{example}

\vspace{1mm}

\begin{example}\label{ñ} {\em Let $V$ be $\L^{2}$ with the origin point removed (Figure
3):
\[
V=\L^{2}\setminus \{(0,0)\}.
\]
The GKP pre-completion $V^{\sharp}$ of this spacetime attaches at
the origin two ideal endpoints given by the pairs $(P,\emptyset)$,
$(\emptyset,F)$. This provides a simple example of the
non-equivalence between the notions of {\em limit} and {\em
endpoint} of a chain: the pair $(\emptyset,F)\in V^{\sharp}$ is a
limit of the chain $\{{\bf i}(p_{n})\}\subset {\bf i}[V]$, however,
the unique endpoint of $\{p_{n}\}$ in $V^{\sharp}$ is instead the
pair $(P,\emptyset)$. Consider now the completion resulting from
replacing in $V^{\sharp}$ the pair $(P,\emptyset)$ by $(P,F)$. In
this case, the sequence constantly equal to $(P,F)$ converges with
the $chr$-topology to both, $(\emptyset,F)$, $(P,F)$, showing that
this topology is not $T_{1}$ for this completion. The (unique)
chronological completion $\overline{V}$ of $V$ only attaches at the
origin the ideal point $(P,F)$, showing in particular that $V$ is
not globally hyperbolic (Theorem \ref{globhyp}). On the other hand,
this example shows that ``boundary'' need not be metrically
infinitely far along geodesics, as illustrated by the curve
$\gamma(t)=(0,t)$, $t<0$.}

\begin{figure}[ht]
\begin{center}
\includegraphics[width=4cm]{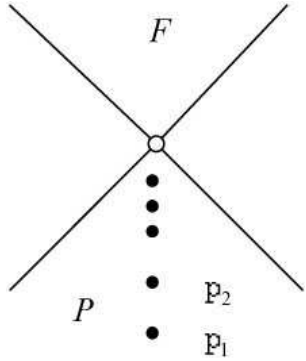}
\caption{Minkowski plane $\L^{2}$ with the origin removed.}
\end{center}
\end{figure}

\end{example}

\vspace{1mm}

\begin{example}\label{1} {\em Let $V$ be $\L^{2}$ with the vertical segment ${\cal
V}_{+}=\{(0,t): t\geq 0\}$ removed (Figure 4):
$$V=\L^{2}\setminus \{(0,t): t\geq 0\}.$$ Let $\overline{V}$ be
the (unique) chronological completion of $V$. The pairs
$(P,F_{l})$, $(P,F_{r})$ are the unique endpoints in
$\overline{V}$ of the chains $\{q_{n}\}$, $\{p_{n}\}$, resp. They
represent two ideal endpoints attached at the extreme of ${\cal
V}_{+}$. These pairs are also endpoints of the future chain
$\{r_{n}\}$, and thus, limits of $\{{\bf i}(r_{n})\}$ (recall
Theorem \ref{compatible} {\it (2)}). Therefore, $\overline{V}$ is
non-Hausdorff with the $chr$-topology. If we extend this analysis
to the whole line ${\cal V}_{+}$, we obtain that $\partial(V)$
contains two copies of ${\cal V}_{+}$, with only the extreme ideal
points $(P,F_{l})$, $(P,F_{r})$ being non-Hausdorff related.

On the other hand, observe that the chain $\{r_{n}\}$ generates the
pair $(P,F_{l}\cup F_{r})\in V_{p}\times V_{f}$. This pair belongs
to the completion $V^{\flat}$ (see Section \ref{chron}), showing in
particular that some completions may contain pairs whose components
are not necessarily indecomposable.}

\begin{figure}[ht]
\begin{center}
\includegraphics[width=4cm]{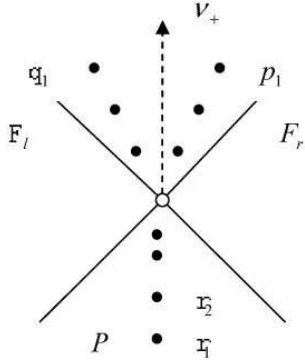}
\caption{Minkowski plane $\L^{2}$ with a vertical segment ${\cal
V}_{+}$ removed.}
\end{center}
\end{figure}

\end{example}

\vspace{1mm}

\begin{example}\label{3} {\em Let $V$ be $\L^{2}$ with the horizontal segment ${\cal
H}_{-}=\{(x,0): x\leq 0\}$ removed (Figure 5):
$$V=\L^{2}\setminus \{(x,0): x\leq 0\}.$$ Let $\overline{V}$ be the (unique) chronological completion
of $V$. For $x<0$, the unique endpoints in $\overline{V}$ of the
chains $\{(x,-1/n)\}$ and $\{(x,1/n)\}$ are the pairs
$(P_{x},\emptyset)$ and $(\emptyset,F_{x})$, resp. However, for
$x=0$ the unique endpoint in $\overline{V}$ of the chains
$\{(0,-1/n)\}$ and $\{(0,1/n)\}$ is the pair $(P_{0},F_{0})$.
Therefore, in this case the chronological completion
$\overline{V}$ contains two copies of ${\cal H}_{-}$ with the
right extreme points of the copies identified via $(P_{0},F_{0})$.
On the other hand, we can ask for the limit of the sequence
$\sigma=\{(P_{x_{n}},\emptyset)\}_{n}\subset \overline{V}$, with
$x_{n}=-1/n$ for all $n$. Surprisingly, $\sigma$ does {\em not}
converge to $(P_{0},F_{0})$ with the $chr$-topology, violating the
common intuition inherited from the natural embedding of this
space into Minkowski.

When we consider a completion different from the chronological
one, the corresponding boundary may be {\em non-closed}. In fact,
take for example the completion $V^{end}$, which contains in
particular the endpoint $(P,F)$, with $P=P_{0}\cup I^{-}(p)$ and
$F=I^{+}(p)$, $p=(-1,1)$. Then, the sequence constantly equal to
$(P,F)$, which is obviously contained in the boundary, converges
to ${\bf i}(p)\in {\bf i}[V]$ with the $chr$-topology.

Finally, consider the future chronological completion $\hat{V}$ of
this spacetime. Apart from the obvious limit $I^{-}(p)$, the
sequence of PIPs $\delta=\{I^{-}(p_{n})\}\subset \hat{V}$,
$p_{n}=(-1+1/n,1)$ for all $n$, also converges to $P_{0}$ with the
$\;\widehat{}\;$-topology. This anomalous limit is due to the fact
that $\hat{V}$ and $\;\widehat{}\;$-topology only retain partial
information about the chronology of $V$. This situation contrasts
with our construction, where the full chronology is taken into
consideration. In fact, under our approach, Theorems \ref{artu},
\ref{princ} and \ref{non-hausdorff} imply that $\{{\bf
i}(p_{n})\}_{n}\subset \overline{V}$ only converges to ${\bf i}(p)$,
as expected. }

\begin{figure}[ht]
\begin{center}
\includegraphics[width=11cm]{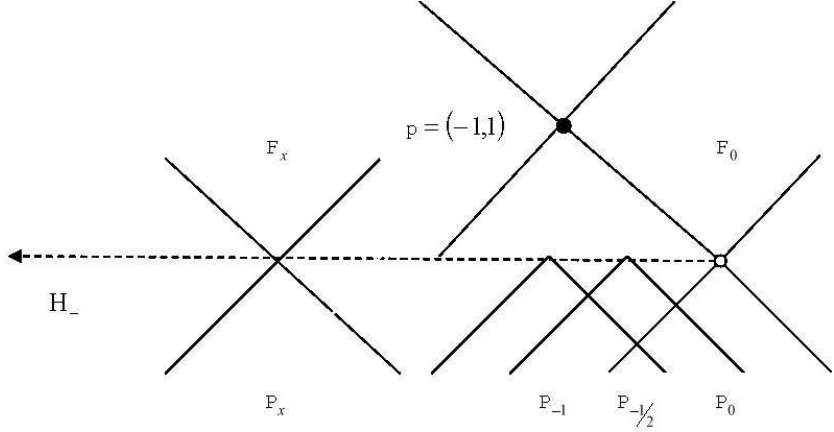}
\caption{Minkowski plane $\L^{2}$ with a horizontal segment ${\cal
H}_{-}$ removed.}
\end{center}
\end{figure}

\end{example}

\vspace{1mm}

\begin{example}\label{5} {\em Consider the spacetime $V$ represented in Figure 6 (this example
comes from \cite[Figure 2]{S1}; see also \cite[Figure 7]{MR}). Here,
infinite null segments $\{L_{n}\}_{n}$ and the point $r$ have been
removed from the Minkowski plane, resulting in a spacetime such that
$I^{-}[\gamma']\varsubsetneq I^{-}[\gamma]$. In this case the
(unique) chronological boundary $\partial(V)$ coincides with the MR
boundary (see \cite{MR}), including the pairs $(I^{-}[\gamma],F)$
and $(I^{-}[\gamma'],\emptyset)$ as endpoints of $\gamma$ and
$\gamma'$, resp. However, the MR topology is different from the
$chr$-topology. In fact, as showed in \cite{MR}, the sequence
constantly equal to $(I^{-}[\gamma],F)$ also converges to
$(I^{-}[\gamma'],\emptyset)$ with the MR topology, and thus, it is
not $T_{1}$. This is not the case for the $chr$-topology, which is
always $T_{1}$ (Theorem \ref{6.4}).}

\begin{figure}[ht]
\begin{center}
\includegraphics[width=4cm]{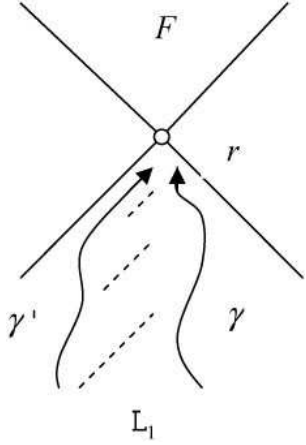}
\caption{Minkowski plane $\L^{2}$ with a sequence of null segments
$\{L_{n}\}_{n}$, including the limit point $r$, removed.}
\end{center}
\end{figure}

\end{example}

\vspace{1mm}

\begin{example}\label{yo} {\em Let $X$ be the disjoint union of $I_{i}=(-\infty,+\infty)_{i}$, $i=1,2$, under the equivalence relation
$0_{1}\sim 0_{2}$ (Figure 7). Endow $X$ with the (quotient of the)
chronology relation given by
\[
x\ll x'\quad\hbox{iff}\quad\left\{\begin{array}{l}
\hbox{either}\quad x<x'\;\;\hbox{and}\;\; x,x'\in I_{i}\;\;
\hbox{for some}\;\; i;
\\ \hbox{or}\quad x<x'\;\;\hbox{and}\;\; 0\geq x\in I_{i},\;\, 0\leq x'\in I_{j}\;\;\hbox{for}\;\; i\neq j.
\end{array}\right.
\]
Then, $X$ becomes a chronological set satisfying
\[
I^{\pm}([0])=I_{1}^{\pm}\cup
I_{2}^{\pm},\qquad\hbox{with}\quad\begin{array}{l}
I_{i}^{-}:=(-\infty,0)_{i}
\\  I_{i}^{+}:=(0,+\infty)_{i}
\end{array}.
\]
Therefore, the past and future of $[0]\in X$ are {\em not}
indecomposable.

\begin{figure}[ht]
\begin{center}
\includegraphics[width=6cm]{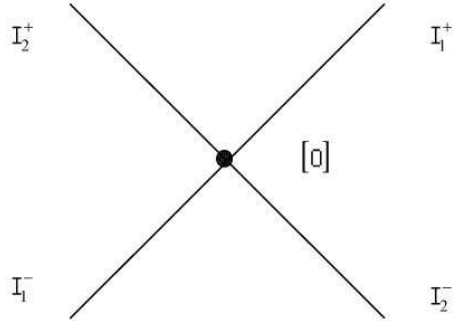}
\caption{Two copies of $\R$ with the zeroes identified.}
\end{center}
\end{figure}

Motivated by this example, consider now the $(2+1)$-spacetime $V$
constructed by deleting from $\L^{3}$ the subsets $\{x=0, -1\leq
t\leq 0\}$ and $\{y=0, 0\leq t\leq 1\}$. In fact, $V$ has two TIPs
$P_{1}$, $P_{2}$ and two TIFs $F_{1}$, $F_{2}$ associated to the
removed origin (Figure 8), which verify $P_{1}\sim_{S} F_{1},F_{2}$
and $P_{2}\sim_{S} F_{1},F_{2}$. Therefore, there are {\em two}
different chronological completions for $V$: one of them attaches at
the origin the ideal endpoints $(P_{1},F_{1})$, $(P_{2},F_{2})$,
while the other one attaches the ideal endpoints $(P_{1},F_{2})$,
$(P_{2},F_{1})$. Both of these chronological completions are {\em
different} from the MR completion, which attaches at the origin
``more'' ideal endpoints: $(P_{1},F_{1})$, $(P_{1},F_{2})$,
$(P_{2},F_{1})$, $(P_{2},F_{2})$. If we additionally delete from
$\L^{3}$ the subset $\{x>0, y>0, t=0\}$ then the MR completion
attaches at the origin the pairs $(P_{1},F_{1})$, $(P_{1},F_{2})$,
$(P_{2},F_{2})$. In this case the spacetime admits an unique
chronological completion, which attaches at the origin the pairs
$(P_{1},F_{1})$, $(P_{2},F_{2})$. Another example of spacetime whose
MR completion also includes spurious ideal points appears in
\cite[Appendix A]{MR}.}

\begin{figure}[ht]
\begin{center}
\includegraphics[width=7cm]{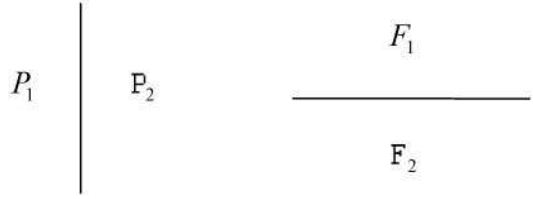}
\caption{Slices showing the cuts made to produce our example, and
the corresponding TIPs and TIFs.}
\end{center}
\end{figure}

\end{example}

\vspace{1mm}

\begin{example}\label{9.5} (Standard Static Spacetimes) {\em The causal boundary for standard static spacetimes has been studied in \cite{H3,FH,AF}. From
the conformal invariance of the causal boundary approach, it is
not a restriction to assume that these spacetimes can be written
as
\[
V=M\times\R,\qquad g=h-dt^{2},
\]
where $(M,h)$ is an arbitrary Riemannian $3$-manifold. In these
spacetimes, the spatial projection $c$ of every inextensible
future-directed timelike curve $\gamma(t)=(c(t),t)$ whose past is
different from the whole spacetime is an {\em asymptotically
ray-like curve}, i.e. an inextensible curve with domain
$[w,\Omega)$, $\Omega\leq\infty$, velocity $|\dot{c}|\leq 1$ and
finite-valued {\em Busemann function} $b_{c}$; that is,
\[
b_{c}:M\rightarrow\R^{*}\equiv\R\cup\{\infty\},\qquad
b_{c}(\cdot):=\lim_{t\rightarrow\Omega}(t-d(\cdot,c(t)))<\infty.
\]
Moreover, the pasts of these curves are totally characterized by
the corresponding Busemann functions $b_{c}$. Therefore, if we
denote by ${\cal B}(M)$ the set of all Busemann functions
associated to asymptotically ray-like curves in $(M,h)$, we obtain
\[
\hat{\partial}(V)={\cal B}(M)\cup\{\infty\}.
\]
In other words, if we define the {\em Busemann boundary} of $(M,h)$
as the quotient space
\[
\partial_{{\cal B}}(M):={\cal B}(M)/\R,
\]
which, in particular, includes the Cauchy boundary
$\partial_{c}(M)$, then the future chronological boundary of $V$
is a cone with apex $i^{+}$ and base $\partial_{{\cal B}}(M)$;
i.e.
\[
\hat{\partial}(V)={\cal B}(M)\cup\{\infty\}\equiv(\partial_{{\cal
B}}(M)\times\R)\cup \{i^{+}\}
\]
Analogously, the past chronological boundary of $V$ is a cone with
apex $i^{-}$ and base $\partial_{{\cal B}}(M)$; i.e.
\[
\check{\partial}(V)=(\partial_{{\cal B}}(M)\times\R)\cup
\{i^{-}\}.
\]

But, what about the (total) chronological boundary? The common
future (resp. common past) of any inextensible future-directed
(resp. past-directed) timelike curve $\gamma$ is non-empty iff
$\gamma$ approaches to some $(t,p)\in\hat{\partial}(V)$ with
$p\in\partial_{c}(M)$; moreover, in this case the common future
(resp. common past) of $\gamma$ coincides with the future (resp.
past) of any past-directed (resp. future-directed) timelike curve
approaching also to $(t,p)\in\check{\partial}(V)$. So, the (total)
chronological boundary is a double cone with base $\partial_{{\cal
B}}(M)$, apexes $i^{+}$, $i^{-}$, and future and past copies of
lines over the same point $p$ of the Cauchy boundary
$\partial_{c}(M)$ identified. Summarizing:
\[
\partial(V)=(\hat{\partial}(V)\cup \check{\partial}(V))/\sim,\qquad\hbox{with}\;\;
(p^{+},t^{+})\sim
(p^{-},t^{-})\;\;\hbox{iff}\;\;\left\{\begin{array}{l}
(p^{+},t^{+})\in \hat{\partial}(V) \\ (p^{-},t^{-})\in
\check{\partial}(V) \\ p^{+}=p^{-}\in\partial_{c}(M) \\
t^{+}=t^{-}\in\R
\end{array}\right.
\]
(see Figure 9).

On the other hand, the $chr$-topology on $\overline{V}$ coincides
with the quotient topology over $\sim$ of the topology generated
by the limits operators $\hat{L}$ and $\check{L}$ on $\hat{V}\cup
\check{V}$.}

\begin{figure}[ht]
\begin{center}
\includegraphics[width=10cm]{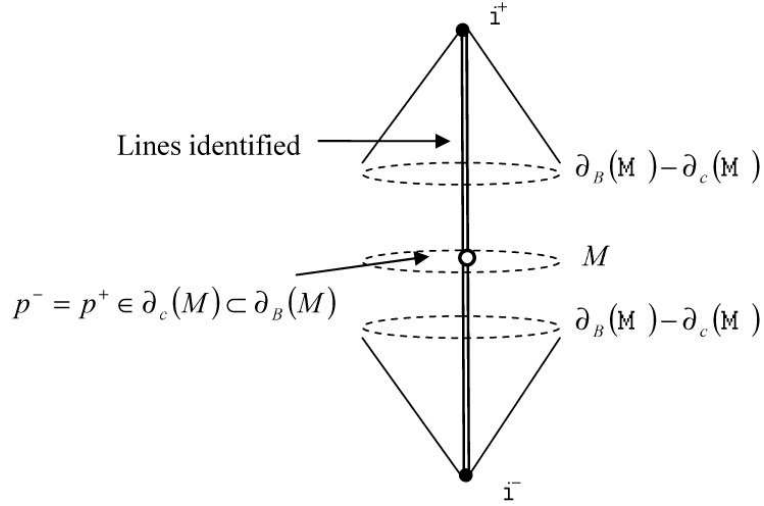}
\caption{Chronological boundary for Standard Static spacetimes.}
\end{center}
\end{figure}

\end{example}

\vspace{1mm}

\begin{example}\label{10} (Locally Symmetric Plane Waves) {\em
Consider $V=\R^{n+2}$ with metric
\[
\langle\cdot,\cdot\rangle
=\langle\cdot,\cdot\rangle_{0}+2du\,dv+H(x)du^{2},
\]
where $\langle\cdot,\cdot\rangle_{0}$ is the canonical metric of
$\R^{n}$ and
\[
H(x)=-\mu_{1}^{2}x_{1}^{2}-\cdots -
\mu_{j}^{2}x_{j}^{2}+\mu_{j+1}^{2}x_{j+1}^{2}+\cdots +
\mu_{n}^{2}x_{n}^{2},\qquad\hbox{with}\;\; j>0
\]
and $\mu_{1}\geq \mu_{2}\geq\cdots\geq \mu_{j}$. The causal boundary
of these spacetimes was first analyzed in \cite{MR0,MR} by using the
MR approach. If we denote by $L^{+}$, $L^{-}$ two copies of the line
$u\in (-\infty,\infty)$, and define
\[
L=(L^{+}\cup L^{-})/\,R,\quad\qquad u\; R\; u'\Longleftrightarrow
u\in L^{+},\; u'\in L^{-},\; u=u'-\pi/\mu_{1},
\]
the authors found that the MR boundary of $V$ can be represented by
the single line $\partial(V)=L\cup\{i^{+},i^{-}\}$. Moreover, in
this case the MR construction coincides with our construction (see
\cite{FS}). Therefore, the chronological boundary of $V$ can be also
represented by this single line (see Figure 10). This picture agrees
with the result previously obtained in \cite{BN} for the maximally
symmetric case by using the conformal approach (see \cite[Section
5]{MR} for a brief discussion).

\begin{figure}[ht]
\begin{center}
\includegraphics[width=5cm]{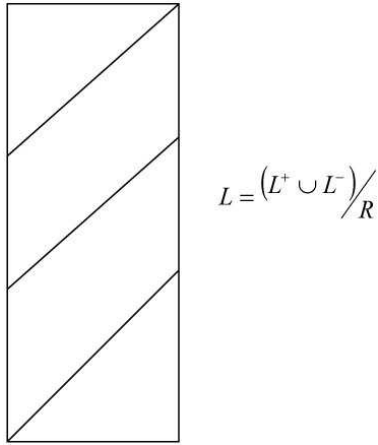}
\caption{Chronological boundary for Locally Symmetric Plane Waves as
pictured in \cite{MR}.}
\end{center}
\end{figure}

This $1$-dimensional character of the boundary admits an intriguing
interpretation in terms of the global causal behavior of the wave.
In fact, in \cite{FS0} the authors found that the causality of {\em
generalized wave type spacetimes}
\begin{equation}\label{genwav}
V=M\times\R^{2},\qquad
\langle\cdot,\cdot\rangle_{z}=\langle\cdot,\cdot\rangle_{x}+2du\,dv+H(x,u)du^{2},
\end{equation}
where $(M,\langle\cdot,\cdot\rangle_{x})$ is a Riemannian
$n$-manifold and $H:M\times\R\rightarrow\R$ a smooth function,
presents a critical behavior with respect to the metric coefficient
$-H$. More precisely, these waves pass from being globally
hyperbolic for $-H$ subquadratic (and $M$ complete) to being
non-distinguishing for $-H$ superquadratic. This gap in the causal
ladder has to do with a sort of degeneracy for the chronology of the
wave ``at infinity'', which, in this case, translates into a {\em
low dimensionality} of the boundary.

We refer the reader to \cite{MR0,HR,FS} for a systematic study (of
increasing generality) of the causal boundary for spacetimes
(\ref{genwav}), including plane waves and pp-waves.}

\end{example}

\section{Acknowledgments}

I would like to thank Professors Steven Harris and Miguel
S\'{a}nchez for useful discussions and comments. Part of this work
was completed during a research stay at Department of Mathematics,
Stony Brook University. Partially supported by MECyD Grant
EX-2002-0612, JA regional Grant P06-FQM-01951 and MEC Grants
RyC-2004-382 and MTM2007-60731.

\end{document}